\documentclass[%
reprint,
pox,
amsmath,amssymb,
aps,
]{revtex4-2}

\pdfoutput=1

\usepackage{graphicx}% Include figure files
\usepackage{dcolumn}% Align table columns on decimal point
\usepackage{bm}% bold math

\usepackage{hyperref}

\begin{document}
	\preprint{APS/123-QED}
	
	\title{Effect of Frequency-Dependent Viscosity on Molecular Friction in Liquids}
	
	\author{Henrik Kiefer}
	\author{Domenico Vitali}
	\author{Benjamin A. Dalton}
	\author{Laura Scalfi}
	\author{Roland R. Netz}
	\email{rnetz@fu-berlin.de}
	\affiliation{%
		Fachbereich Physik, Freie  Universität  Berlin,  Arnimallee  14,  14195  Berlin,  Germany
	}%
	
	\date{\today}

	\begin{abstract}
		The relation between the frequency-dependent friction of a molecule in a liquid
		and the hydrodynamic properties of the liquid is fundamental for molecular dynamics.
		We investigate this connection for a water molecule moving in liquid water using all-atomistic molecular dynamics simulations and linear hydrodynamic theory. We analytically calculate the frequency-dependent friction of a sphere with finite surface slip moving in a viscoelastic compressible fluid by solving the linear transient Stokes equation, including frequency-dependent shear and volume viscosities, both determined from MD simulations of bulk liquid water. We also determine the frequency-dependent friction of a single water molecule moving in liquid water, as defined by the generalized Langevin equation from MD simulation trajectories. By fitting the effective sphere radius and the slip length, the frequency-dependent friction and velocity autocorrelation function from the transient Stokes equation and simulations quantitatively agree. This shows that the transient Stokes equation accurately describes the important features of the frequency-dependent friction of a single water molecule in liquid water and thus applies down to molecular length and time scales, provided accurate frequency-dependent viscosities are used. The frequency dependence of the shear viscosity of liquid water requires careful consideration of hydrodynamic finite-size effects to observe the asymptotic hydrodynamic power-law tail. In contrast, for a methane molecule moving in water, the frequency-dependent friction cannot be predicted based on a homogeneous model, which suggests, supported by the extraction of a frequency-dependent surface-slip profile, that a methane molecule is surrounded by a finite-thickness hydration layer with viscoelastic properties that are significantly different from bulk water.
		
		\begin{description}
			\item[Subject Areas]
			Soft Matter,
			Statistical Physics, Fluid Dynamics, Complex Systems
		\end{description}
	\end{abstract}
	
	\maketitle
	
	\newpage
	
	\section{Introduction}
	The friction force acting on a solute molecule in a liquid environment exhibits a delayed non-Markovian response due to the finite relaxation 
	time of the solvating liquid degrees of freedom   \cite{mori1965transport, zwanzig1965time,carof2014two,lesnicki2016molecular,Ayaz2022}. 
	Such memory effects occur on time scales that range between sub-picoseconds up to microseconds and even seconds,
	depending on the type and complexity of the system \cite{jung2017iterative, jung2017frequency, daldrop2017external, kowalik2019memory,mitterwallner2020non}. Including time- or frequency-dependent friction in an appropriate theoretical framework allows for the accurate modeling of macromolecular dynamics
	in liquid environments  \cite{daldrop2018butane,ayazproteins, dalton2022protein,dalton2024role} and for the proper viscoelastic description of soft matter \cite{sollich1997rheology, mizuno2007nonequilibrium, winter2012active,hiraiwa2018systematic}.
	A fundamental question concerns the connection between the macroscopic hydrodynamic equations and the molecular friction acting on a particle or a molecule in a fluid. 
	Numerous studies investigated this connection by comparing the frequency-dependent friction acting on a particle, as described by the generalized Langevin equation, with the friction obtained by solving the hydrodynamic equations for the fluid flow around 
	a spherical particle
	\cite{chow1972effect,chow1973brownian,hauge1973fluctuating,bedeaux1974generalization,hinch1975application,espanol1995propagation,felderhof2005backtracking,chakraborty2011velocity,tatsumi2012direct,mizuta2019bridging}. 
	Pioneering work in this direction was done by Zwanzig \textit{et al.} \cite{zwanzig1970hydrodynamic} and later by Metiu \textit{et al.} \cite{metiu1977hydrodynamic}, 
	who obtained the time- or frequency-dependent friction by solving the linearized Stokes equation for a spherical particle in the presence
	of a frequency-dependent shear viscosity, described by a Maxwell model with a single relaxation time. \\
	\indent Since a considerable discrepancy between the friction obtained from the solution of the Stokes equation 
	and the friction derived from 
	the velocity autocorrelations obtained in molecular dynamics simulations was found, 
	especially at high frequencies, it was concluded that hydrodynamic theory does not work on molecular time and length scales
	\cite{zwanzig1970hydrodynamic,metiu1977hydrodynamic}.
	Such a breakdown of hydrodynamics would most plausibly be explained by spatial non-locality in the fluidic viscous response, which in principle could be treated in reciprocal space but would render the Stokes solution for the frequency-dependent friction around a sphere invalid. 
	However, a critical limitation in the comparison in  \cite{zwanzig1970hydrodynamic,metiu1977hydrodynamic}
	is that a Maxwell model with a single relaxation time was used for the shear viscosity in the solution of the Stokes equation.
	In fact, viscosity spectra measured in experiments \cite{slie1966ultrasonic,pelton2013viscoelastic} and extracted from molecular dynamics (MD) simulations of water \cite{medina2011molecular,fanourgakis2012determining,schulz2020molecular}, 
	indicate pronounced deviations from a simple Maxwell model, especially at high frequencies in the THz regime. 
	This is the frequency range where deviations between the friction from hydrodynamic predictions and molecular simulation were found, so it is unclear whether spatially homogeneous hydrodynamic theory breaks down or whether an inappropriate model
	for the shear viscosity was used. \\
	\indent In the present work, we reconsider the connection between macroscopic hydrodynamics and molecular friction; for this, we consider a single water molecule in a liquid water environment. 
	We first analytically calculate the frequency-dependent friction acting on a sphere using the linearized Stokes equation 
	in the presence of frequency-dependent shear and volume viscosities, finite compressibility, and a finite surface slip
	\cite{erbacs2010viscous}. In contrast to previous work \cite{zwanzig1970hydrodynamic,metiu1977hydrodynamic,felderhof2005backtracking,chakraborty2011velocity} we do not use a phenomenological Maxwell model for the viscosities but rather employ frequency-dependent shear and volume viscosities extracted from MD simulations. 
	We, in detail, investigate the influence of compressibility and the frequency dependence of the volume viscosity on the friction function at high frequencies. 
	We finally compare the friction calculated from the transient Stokes equation with the friction extracted from MD simulations 
	using the framework of the generalized Langevin equation. \\
	\indent Using the surface-slip parameter and the sphere radius that appear in the hydrodynamic prediction of the friction as free fit parameters, we 
	find that the frequency-dependent friction of a water molecule extracted directly from MD simulations is in good agreement with the 
	hydrodynamic predictions for frequencies up to 10 THz. \\
	\indent This establishes the long-sought link between macroscopic hydrodynamics and the friction acting on a molecule in a fluid
	and shows that the continuum hydrodynamic equations work for water down to the scale of a single water molecule. 
	It turns out that the macroscopic shear viscosity of water shows pronounced multi-modal behavior as a function of frequency and
	thus cannot be described by a Maxwell model with a single relaxation time, which explains why previous 
	attempts to derive the frequency-dependent molecular friction from hydrodynamic theory failed.
	The fitted hydrodynamic radius and slip length obtained from our comparison agree with recent results extracted from
	the translational and rotational diffusivities of a water molecule in liquid water\cite{zendehroud_slip}, 
	which demonstrates that our approach is physically sound. \\
	\indent The friction calculated from the Stokes equation exhibits a power-law tail for long times, attributed to the so-called  Basset-Boussinesq force \cite{boussinesq1903theorie,alder1970decay,hansen1990theory}.
	However, on the time scales reachable with MD  simulations of water, 
	the long time tail, which corresponds to a negative force, is completely dominated
	by the frequency-dependent shear viscosity, which produces positive friction up to times of a few picoseconds, and is for longer times masked by finite-size effects \cite{yeh2004system,asta2017transient, scalfi_pbc},
	in perfect agreement between our hydrodynamic predictions and the MD simulation results. 
	Our findings are supported by a comparison of the velocity autocorrelation function computed from the MD simulation and from the hydrodynamic friction including frequency-dependent viscosities and finite compressibility.\\
	\indent We thus find that the macroscopic hydrodynamic equations work surprisingly well down to molecular time and spatial
	scales for homogeneous water, i.e., if one considers the motion of a single water molecule embedded in liquid water,
	if and only if the frequency-dependent shear viscosity of water is used.
	This shows that a wave-vector-dependent viscosity, which would appear in a formally exact formulation
	of the linear-response stress-strain relation, is not necessary. \\ 
	\indent In contrast,
	the friction of a single methane molecule in liquid water is not well described by hydrodynamic theory using liquid water shear and volume viscosities. A reasonable extension of the theory is the introduction of a frequency-dependent slip coefficient. We, therefore, calculate the slip profile which leads to a perfect accordance between the friction of the MD simulation and hydrodynamic theory. The surface-slip profiles of the water and methane simulation include frequency ranges where they are negative, indicating that the local viscosity around the sphere must deviate from the macroscopic shear viscosity.
	It is known that methane is in the water surrounded by a clathrate-like highly-ordered structure.
	We thus conclude that for inhomogeneous liquids, i.e., for the motion of a molecule that differs from the surrounding liquid,
	a homogeneous hydrodynamic model has to be generalized to account
	for the possibly modified viscosity of the solvation layer around a moving host molecule, as recently found for nanoscopic tracer beads moving in hydrogels \cite{schmidt2024nanoscopic}. 
	
	\section{Theory}
	\subsection{Frequency-dependent friction of a sphere from hydrodynamic theory}
	The frequency-dependent friction of a sphere, $\Tilde{\Gamma}^{hyd}(\omega)$, is a complex-valued function that describes the fluid stress response to a small, oscillatory sphere motion and is defined as
	\begin{equation}
		\label{eq:force_hydro}
		\Tilde{F}_i(\omega) =    \delta_{ij}\Tilde{\Gamma}^{hyd}(\omega)  \Tilde{v}_j(\omega),  
	\end{equation}
	where $\Tilde{v}_j(\omega)$ is the frequency-dependent velocity of the sphere and $\Tilde{F}_i(\omega)$ is an external
	force acting on the sphere with radius $a$, and the indices $i,j \in \{x,y,z\}$.
	In our work, we define the spatial and temporal Fourier transformation (FT) of a function as 
	$\Tilde{f}(\Vec{k},\omega)    =   \int dt\:d^3r \:  f(\Vec{r},t)   e^{- i(k_ir_i + \omega t)}$.
	The friction of a sphere in a liquid can be derived from the Navier-Stokes equation, which originates from local momentum conservation
	\cite{stokes1851effect,batchelor2000introduction,kim2013microhydrodynamics}
	and follows as 
	\begin{eqnarray}
		\label{eq:resp_sphere} \nonumber
		\Tilde{\Gamma}^{hyd}(\omega) = &&  \frac{4\pi\Tilde{\eta}(\omega) a}{3}W^{-1} \Bigl\{(1+\hat{\lambda})(9+9\hat{\alpha}+\hat{\alpha}^2)(1+2\hat{b}) \\ && + (1+\hat{\alpha})[2\hat{\lambda}^2(1+2\hat{b})+\hat{b}\hat{\alpha}^2(1+\hat{\lambda})]\Bigr\},
	\end{eqnarray}
	where $W$ is given by
	\begin{equation}
		\label{eq:W}
		W = (2+2\hat{\lambda}+\hat{\lambda}^2)(1+\hat{b}(3+\hat{\alpha}))+(1+\hat{\alpha})(1+2\hat{b})\hat{\lambda}^2/\hat{\alpha}^2.
	\end{equation}
	Finite slip at the spherical surface is described by the dimensionless slip length, $\hat{b} = b/a$. 
	The dimensionless decay constants $\hat{\alpha} = a\alpha$ and $\hat{\lambda} = a\lambda$ are defined by
	\begin{equation}
		\label{eq:alpha}
		\alpha^2(\omega) = i \omega\rho_0/\Tilde{\eta}(\omega)
	\end{equation}
	and
	\begin{equation}
		\label{eq:lambda}
		\lambda^2(\omega) = \frac{i \omega\rho_0}{4\Tilde{\eta}(\omega)/3 + \Tilde{\zeta}(\omega) - i\rho_0c^2/\omega},
	\end{equation}
	where $c$ is the speed of sound and $\rho_0$ is the mean fluid mass density. 
	The full derivation of Eq.~\eqref{eq:resp_sphere} is presented in Appendix \ref{app:friction_deriv}.\\
	\indent The shear viscosity $\Tilde{\eta}(\omega)$ and volume viscosity $\Tilde{\zeta}(\omega)$ are defined by the stress tensor in the Navier-Stokes equation (Appendix \ref{app:friction_deriv}). If the viscous response decays on length- and time scales that are small compared to those on which the velocity gradients of the fluid $\nabla_j v_i(\Vec{r},t)$ varies, one can approximate the viscous response as frequency- and momentum-independent, which defines a Newtonian fluid. However, we will explicitly consider frequency-dependent viscosities in this work. The shear viscosity $\eta(t)$ is calculated from the autocorrelation of the trace-less stress tensor.
	On the other hand, the volume viscosity $\zeta(t)$ quantifies a fluid's  dissipative
	response to compression \cite{landau1959fluid} and is crucial for describing processes such as sound propagation or shock waves \cite{fanourgakis2012determining}, it can be calculated from the autocorrelation of instantaneous pressure fluctuations.
	Often, the volume viscosity is neglected, which corresponds to the Stokes hypothesis \cite{stokes1880theories, jaeger2018bulk}. 
	However, previous simulations and experiments found that the volume viscosity for water is non-negligible and can even be larger than the shear viscosity \cite{harris2004temperature}. Thus, we explicitly consider a non-vanishing volume viscosity and will carefully examine its influence on the 
	frequency-dependent friction. 
	
	\begin{figure*}
		\centering
		\includegraphics[width=1
		\linewidth]{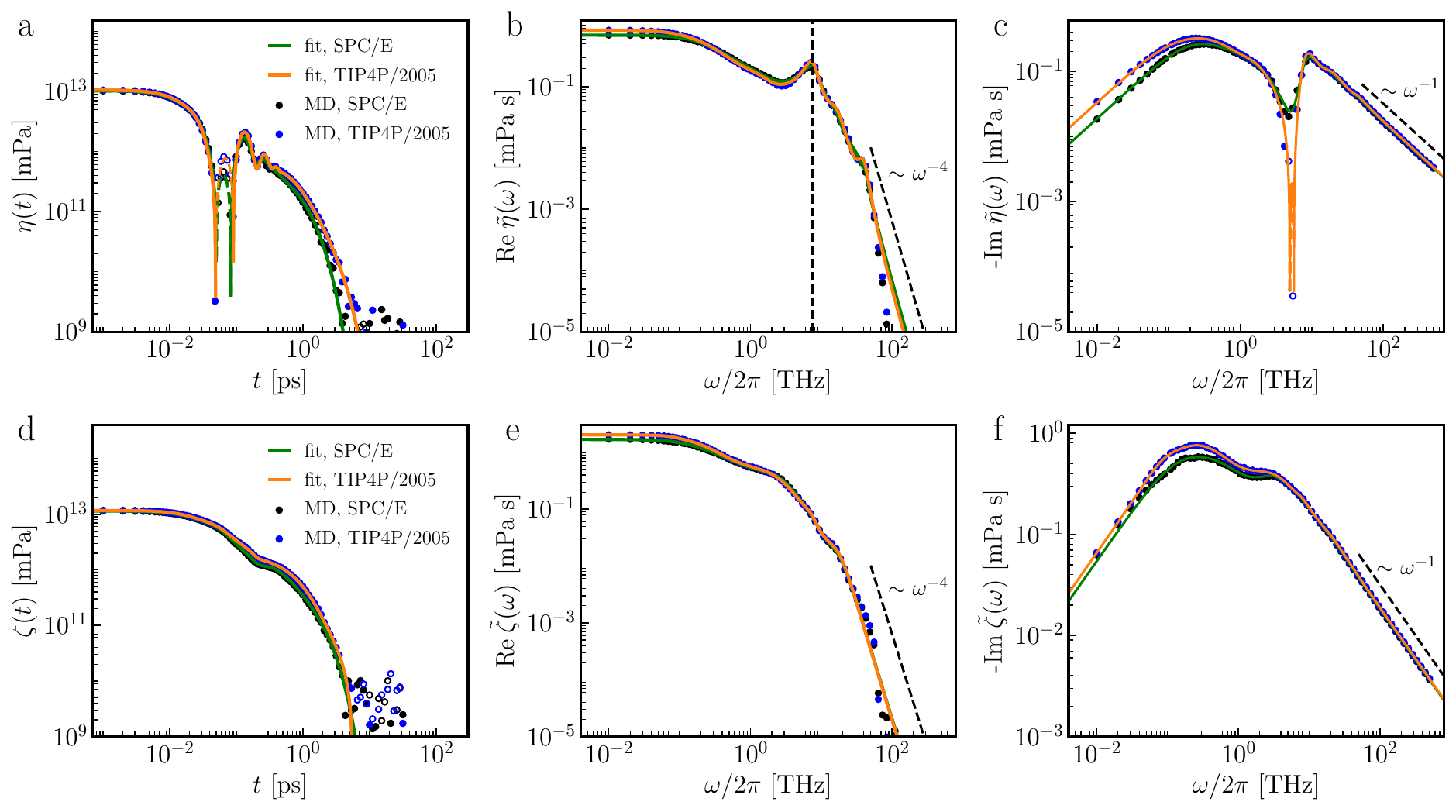}
		\caption{Extracted shear viscosity $\eta$ and volume viscosity $\zeta$ from MD simulations of SPC/E and TIP4P/2005 water.  (a) Shear viscosity in the time domain (circles) with fits (lines) according to Eq.~\eqref{eq:visc_model_time} and the fitting parameters in Table~\ref{tab:fit_visc} (Appendix \ref{app:fit_params}). (b, c) Real and imaginary parts of the shear viscosity in the frequency domain with fits according to Eq.~\eqref{eq:visc_model}. The vertical dashed line denotes the resonance frequency $f_{r,III} \approx 7.11\:\text{THz}$ 
			from the viscoelastic exponential-oscillatory model. (d) Volume viscosity in the time domain (circles) with fits (lines) according to Eq.~\eqref{eq:bulk_model_time} and the fitting parameters in Table~\ref{tab:fit_bulk_visc} (Appendix \ref{app:fit_params}). (e, f) Real and imaginary parts of the volume viscosity in the frequency domain with fits according to Eq.~\eqref{eq:bulk_model}. Dashed lines and empty circles denote negative values. Note that all data is shown in logarithmic spacing for better visibility.}
		\label{fig:visc_water}
	\end{figure*}
	
	\subsection{Frequency-dependent particle friction from the generalized Langevin equation}
	For a particle with mass $m$, the dynamics can be described by the generalized Langevin equation (GLE) \cite{mori1965transport,zwanzig1965time,kowalik2019memory}
	\begin{equation}
		\label{eq:GLE}
		m\Ddot{\Vec{r}}(t) = -\Vec{\nabla} U[\Vec{r}(t)] - \int_{-\infty}^t dt'\:\Gamma(t-t') \Dot{\Vec{r}}(t') + \Vec{F}^R(t),
	\end{equation}
	where $-\nabla U[\Vec{r}(t)$] is the force due to a potential, $\Gamma(t)$ the friction function, often called the memory kernel, and $\Vec{F}_R(t)$ the random force, which has zero mean and a variance of
	\begin{equation}
		\label{eq:FDT}
		\langle F^R_i (t) F^R_j(t')\rangle = k_B T \delta_{ij} \:\Gamma(|t-t'|),
	\end{equation}
	where  $k_BT$ is the thermal energy.
	The stationary friction coefficient of  the particle  $\gamma_0$ is determined by the integral over the memory kernel, 
	i.e., $\gamma_0 = \int_0^\infty \Gamma(t) dt$.
	We assume isotropic fluids and thus consider only the $x$-component of the particle position. 
	In the absence of a potential, $ U = 0$, 
	the solution of the GLE  in Eq.~\eqref{eq:GLE} in Fourier space reads for the particle velocity along $x$ as
	\begin{align}
		\label{eq:GLE_FT}
		\Tilde{v}_x (\omega) = \frac{\Tilde{F}_R(\omega)}{\Tilde{\Gamma}_+ (\omega) + i\omega m},
	\end{align}
	where we use the single-sided memory function $\Gamma_+(t) = \Gamma(t)$ for $t\geq0$ and $\Gamma_+(t) = 0$ for $t<$ 0. 
	In Appendix \ref{app:momentum_sphere}, we show by calculating the fluid momentum outside a moving sphere from the transient Stokes equation that the hydrodynamic friction $\Tilde{\Gamma}^{hyd}(\omega)$ in Eq.~\eqref{eq:resp_sphere} does not include inertial effects inside the sphere. Thus, by comparing Eqs.~\eqref{eq:force_hydro} and Eq.~\eqref{eq:GLE_FT} we conclude that $\Tilde{\Gamma}^{hyd}(\omega)$ and $\Tilde{\Gamma}_+(\omega)$ describe the same quantity, namely the frequency-dependent friction of a particle due to the surrounding fluid.
	
	\section{Results and Discussion}
	
	\subsection{Frequency-dependent shear and volume viscosities}
	\label{sec:visc_freq}
	\begin{figure*}
		\centering
		\includegraphics[width=0.98
		\linewidth]{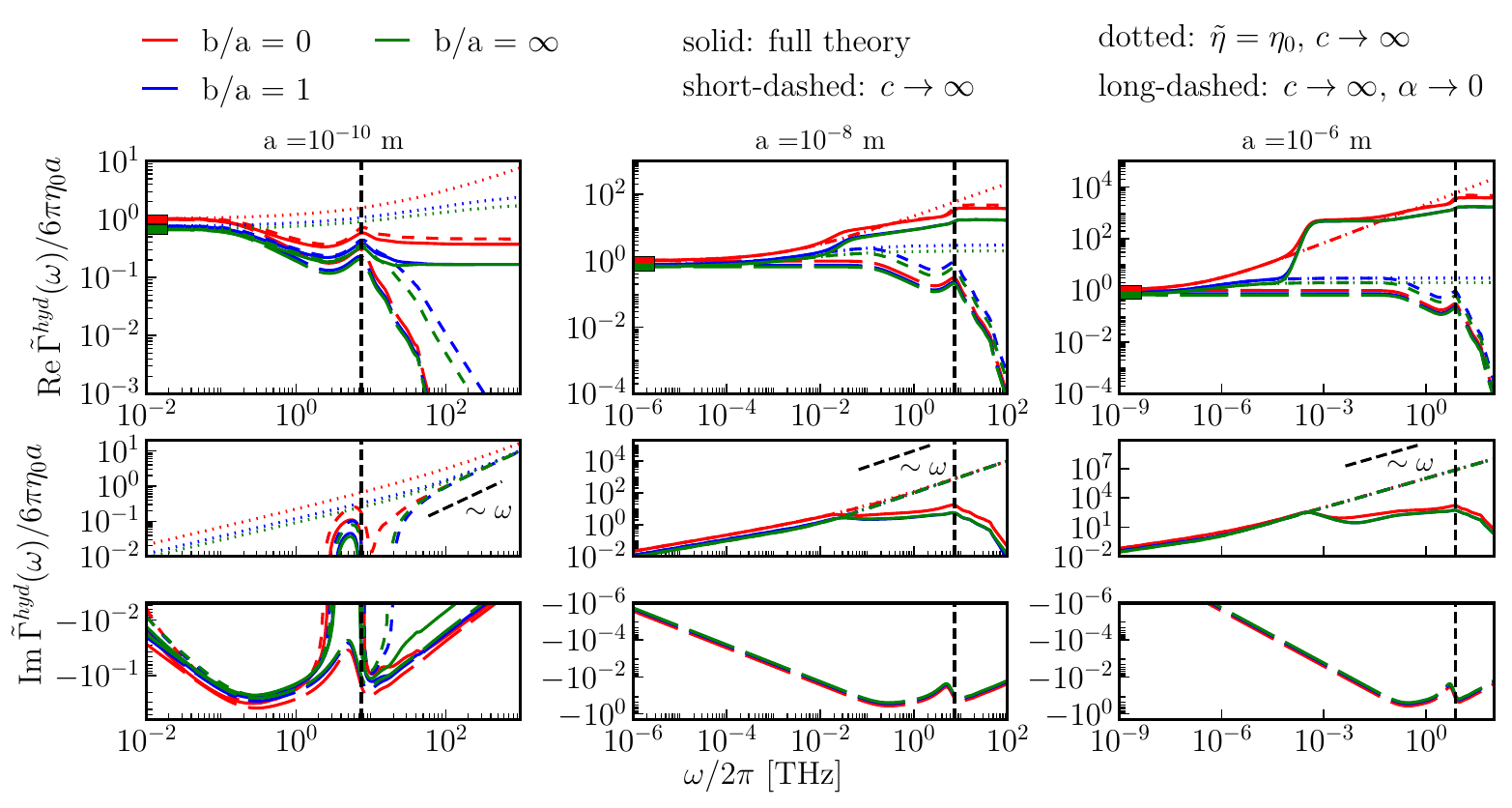}
		\caption{Real and imaginary parts of the frequency-dependent rescaled friction of a sphere $\Tilde{\Gamma}^{hyd}(\omega) = \:\text{Re}\:\Tilde{\Gamma}^{hyd}(\omega) + i\:\text{Im}\:\Tilde{\Gamma}^{hyd}(\omega)$, given by Eq.~\eqref{eq:resp_sphere}, for various slip parameters $\hat{b} = b/a$ and sphere radii $a$. Here we use the viscoelastic model for the shear viscosity $\Tilde{\eta}(\omega)$ in Eq.~\eqref{eq:visc_model} and volume viscosity $\Tilde{\zeta}(\omega)$ in Eq.~\eqref{eq:bulk_model} for SPC/E water from Fig.~\ref{fig:visc_water}. The results are normalized by the steady-state viscosity  $\eta_0 = \sum_{j} \eta_{0,j} = 0.70$ mPa s. We set the water density to $\rho_0$ = 10$^3$ kg/m$^3$ and the speed of sound to $c=1.51 \cdot 10^3\:$m/s \cite{sedlmeier2014charge}. 
			The red and green bars to the left denote the $\omega \to 0$ limits \cite{stokes1851effect}, $6\pi\eta_0 a$ for $\hat{b} \to 0$ and $4\pi\eta_0 a$ for $\hat{b} \to \infty$, respectively. 
			We compare the results of the full hydrodynamic theory (solid lines) with the approximation $c\to \infty$ (short-dashed lines), which represents the incompressible limit, with the double approximation $c \to \infty$ and $\Tilde{\eta} = \eta_0$ (dotted lines), and with the double approximation $c\to \infty$ and $\alpha \to 0$ (long-dashed lines), which represents the generalized Stokes-Einstein relation (GSER, shown in Eq.~\eqref{eq:gser}).
			The vertical dashed line denotes the resonance frequency $f_{r,III} \approx 7.11\:\text{THz}$
			from the viscoelastic exponential-oscillatory model in Fig.~\ref{fig:visc_water}.}
		\label{fig:resp_fct_visc_freq}
	\end{figure*}
	
	\begin{figure*}
		\centering
		\includegraphics[width=0.98
		\linewidth]{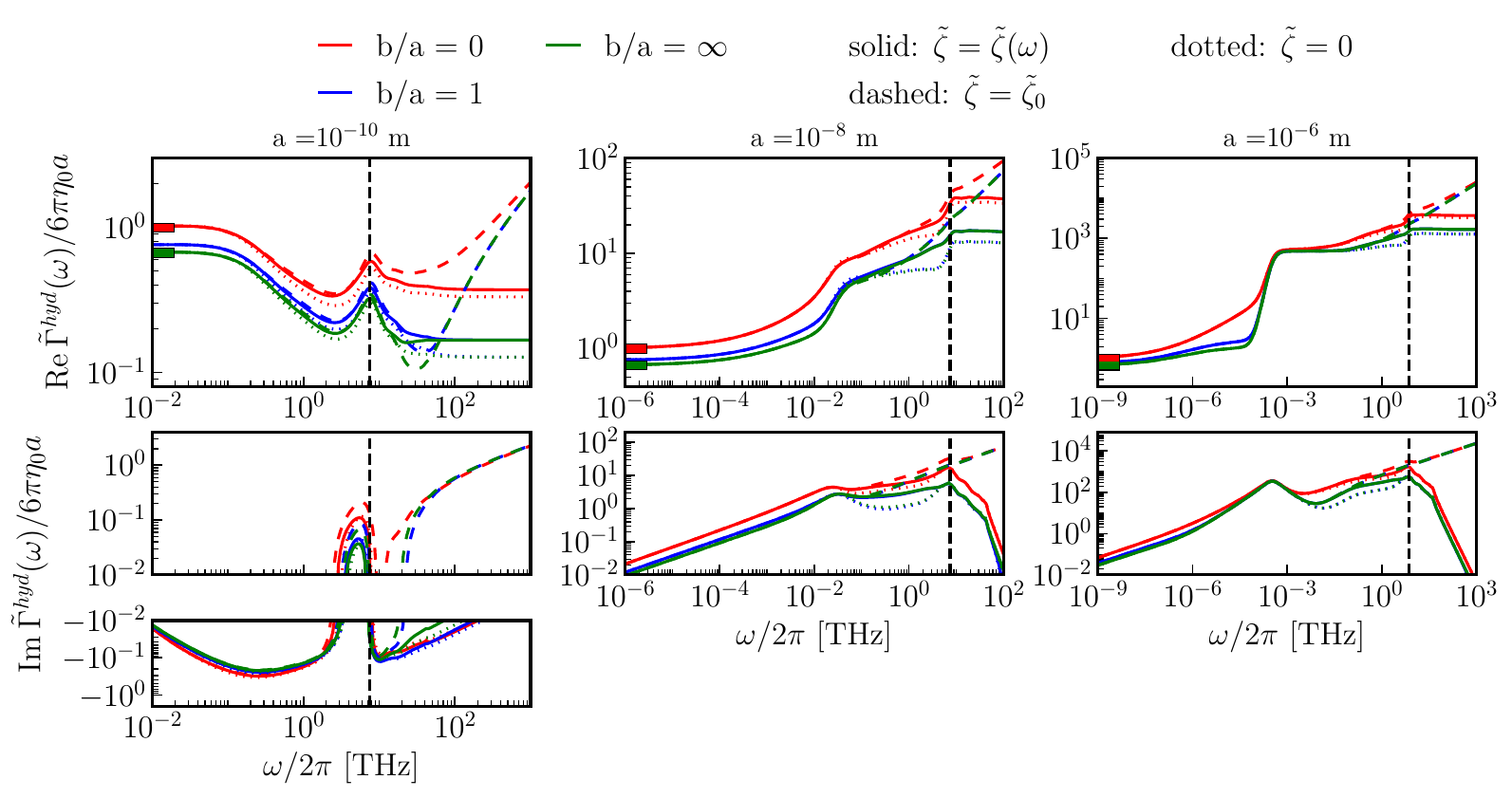}
		\caption{Investigation of the volume viscosity's influence on the hydrodynamic friction of a sphere $\Tilde{\Gamma}^{hyd}(\omega) = \:\text{Re}\:\Tilde{\Gamma}^{hyd}(\omega) + i\:\text{Im}\:\Tilde{\Gamma}^{hyd}(\omega)$, given by Eq.~\eqref{eq:resp_sphere}, for various slip parameters $\hat{b} = b/a$ and sphere radiii $a$. Here we use the viscoelastic model for the shear viscosity $\Tilde{\eta}(\omega)$ in Eq.~\eqref{eq:visc_model} for SPC/E water from Fig.~\ref{fig:visc_water}. The results are normalized by the steady-state viscosity  $\eta_0 = \sum_{j} \eta_{0,j} = 0.70$ mPa s. We set the water density to $\rho_0$ = 10$^3$ kg/m$^3$ and the speed of sound to $c=1.51 \cdot 10^3\:$m/s \cite{sedlmeier2014charge}. We show results for frequency-independent volume viscosity $\zeta_0 = \sum_{j} \zeta_{0,j} = 1.69$ mPa s (dashed lines), for vanishing volume viscosity ($\Tilde{\zeta} = 0$, dotted lines) and for the full frequency-dependent volume viscosity from in Fig.~\ref{fig:visc_water}(d) (solid lines). The red and green bars denote the $\omega \to 0$ limits \cite{stokes1851effect}, $6\pi\eta_0 a$ for $\hat{b} \to 0$ and $4\pi\eta_0 a$ for $\hat{b} \to \infty$, respectively. The vertical dashed line denotes the resonance frequency $f_{r,III} \approx 7.11\:\text{THz}$ 
			from the viscoelastic exponential-oscillatory model in Fig.~\ref{fig:visc_water}.}
		\label{fig:resp_fct_visc_freq_volume_visc}
	\end{figure*}
	
	We analyze MD simulations at temperature $T=300$ K for the SPC/E and TIP4P/2005 water models (see Appendix \ref{app:MD_sim} for simulation details), and compute the frequency-dependent shear $\Tilde{\eta}(\omega)$ and volume viscosity $\Tilde{\zeta}(\omega)$ (see Appendix \ref{app:green_kubo}).
	The Newtonian fluid, as defined in Eq.~\eqref{eq:stress_tensor_nofreq} in Appendix \ref{app:friction_deriv}, is a standard model to describe large-scale and long-time hydrodynamics of liquid water \cite{lifshitz1987course,acheson1991elementary,batchelor2000introduction}. However, in earlier experimental investigations and MD simulations, it was found that at high frequencies, typically in the THz regime, liquid water deviates from the Newtonian fluid model \cite{slie1966ultrasonic,pelton2013viscoelastic,omelyan2005wavevector,lacevic2016viscoelasticity,o2019viscoelasticity,bocquet2010nanofluidics,ruijgrok2012damping,cunsolo1999experimental,carey1998measurement,carlson2020exploring,schulz2018collective}
	and that the shear viscosity decreases at high frequencies \cite{schulz2020molecular}. \\
	\indent In Fig.~\ref{fig:visc_water}(a, b, c), we show the extracted shear viscosity in the time and frequency domain from both water models.
	The TIP4P/2005 model spectra are very similar to the SPC/E model; both exhibit a pronounced peak in the real and imaginary parts around 7-8 THz. The value of the steady-state shear viscosity for the SPC/E model of $\eta_0 = \sum_{j} \eta_{0,j} = 0.70$ mPa s is lower than the value $\eta_0 = 0.84$ mPa s for the TIP4P/2005 model \cite{fanourgakis2012determining,tazi2012diffusion, kowalik2019memory}. 
	We fit the shear viscosity with a sum of six exponential-oscillating functions \cite{schulz2020molecular}
	\begin{eqnarray}
		\label{eq:visc_model_time} \nonumber
		\eta(t) = &&\Theta(t)\Bigl\{\sum_{j=I}^{VI} \frac{\eta_{0,j}\tau_{n,j}}{\tau_{o,j}^2}e^{-t/2\tau_{n,j}}\Bigl[\frac{1}{\kappa_j}\sin{\Bigl(\frac{\kappa_j}{2\tau_{n,j}}t\Bigr)} \\ && + \cos{\Bigl(\frac{\kappa_j}{2\tau_{n,j}}t\Bigr)}\Bigr]\Bigr\},
	\end{eqnarray}
	where $\kappa_j = \sqrt{4(\tau_{n,j}/\tau_{o,j})^2-1}$,
	which in the frequency domain reads as
	\begin{equation}
		\label{eq:visc_model}
		\Tilde{\eta}(\omega) =  \sum_{j=I}^{VI} \eta_{0,j}\frac{1+i\omega\tau_{n,j}}{1+i\omega\tau_{o,j}^2/\tau_{n,j}-\omega^2\tau_{o,j}^2},
	\end{equation}
	as described in Appendix \ref{app:fitting_procedure}. Depending on the value of 
	$\kappa$, a viscosity component displays a single-exponential decay with
	oscillations (finite real part of $\kappa$), or is a sum of two non-oscillating exponentials (imaginary $\kappa$).
	We find the fit function for both water models to perfectly describe the MD data in Fig.~\ref{fig:visc_water}(a, b, c). We provide the fit parameters in Table~\ref{tab:fit_visc} in Appendix \ref{app:fit_params}, and the individual components for the SPC/E water shear viscosity in Appendix \ref{app:fit_params_plot_spce}. 
	Our shear viscosity model decomposes the viscosity spectrum into modes that describe distinct dynamical processes \cite{schulz2020molecular}.
	The oscillation component I is  due to hydrogen-bond network topology changes, i.e., changes of nearest-neighbor water pairs,
	while component II is due to hydrogen-bond stretch vibrations of water pairs, they are both overdamped \cite{schulz2020molecular}.
	The large peak around 7 THz due to 
	the exponential-oscillatory component III  describes  vibrations of hydrogen-bonded water pairs 
	and agrees in position with infrared spectroscopy simulation studies \cite{gonzalez2011flexible}.
	We identify the remaining high-frequency modes IV, V, and VI by comparison with absorption spectra of simulated water 
	\cite{carlson2020exploring} as librational modes, i.e., rotational vibrations of individual water molecules, the splitting is related to the fact that rotations between the three main water axes are not equivalent. 
	\\ \indent In Fig.~\ref{fig:visc_water}(d, e, f), we show the volume viscosity extracted from the MD data, which we determine from the autocorrelation of system pressure fluctuations. The real part of these spectra exhibits no distinct peak in the THz regime, contrary to the shear viscosity, and in agreement with previous simulation results \cite{fanourgakis2012determining}. We fit the volume viscosity data with a sum of five exponential-oscillatory components
	\begin{eqnarray}
		\label{eq:bulk_model_time} \nonumber
		\zeta(t) =&& \Theta(t)\Bigl\{\sum_{j=I}^{V} \frac{\zeta_{0,j}\tau_{v,j}}{\tau_{w,j}^2}e^{-t/2\tau_{v,j}}\Bigl[\frac{1}{\kappa_j}\sin{\Bigl(\frac{\kappa_j}{2\tau_{v,j}}t\Bigr)} \\ && + \cos{\Bigl(\frac{\kappa_j}{2\tau_{v,j}}t\Bigr)}\Bigr]\Bigr\},
	\end{eqnarray}
	where $\kappa_j = \sqrt{4(\tau_{v,j}/\tau_{w,j})^2-1}$. The total complex volume viscosity in the frequency domain is given by
	\begin{equation}
		\label{eq:bulk_model}
		\Tilde{\zeta}(\omega) =  \sum_{j=I}^{V} \zeta_{0,j}\frac{1+i\omega\tau_{v,j}}{1+i\omega\tau_{w,j}^2/\tau_{v,j}-\omega^2\tau_{w,j}^2}.
	\end{equation}
	
	\noindent We find a steady-state value of $\zeta_0 = \sum_{j} \zeta_{0,j} = 1.69$ mPa s for the SPC/E, and 2.04 mPa s for the TIP4P/2005 model, slightly lower than the experimental value (2.4 mPa s for 298 K \cite{harris2004temperature}) and comparable to results from previous MD simulations \cite{fanourgakis2012determining},
	which yielded
	for TIP4P/2005 water  $\zeta_0 = 2.07$ mPa s at 298 K and $\zeta_0 = 2.01$ mPa s at 303 K, 
	and for SPC/E water $\zeta_0 = 1.57$ mPa s at 298 K and $\zeta_0 = 1.45$ mPa s at 303 K.
	\\ \indent For high-frequencies around 100 THz, the fitting functions' real part of both viscosity spectra deviate from the MD data, where according to the models in Eq.~\eqref{eq:visc_model} and \eqref{eq:bulk_model} the real part scales with $\sim \omega^{-4}$ (see dashed black lines). These discrepancies are not a simulation time step issue (Appendix \ref{app:memkernel_time_res}), but due to short comings of the fitting function \cite{straube2020rapid} and occur at frequencies where molecular vibration of real water happens \cite{carlson2020exploring} which are not included in our rigid water models and are not of concern in this work.
	
	\subsection{Frequency-dependent friction of a sphere from hydrodynamic theory including frequency-dependent viscosities}
	\begin{figure*}
		\centering
		\includegraphics[width=1
		\linewidth]{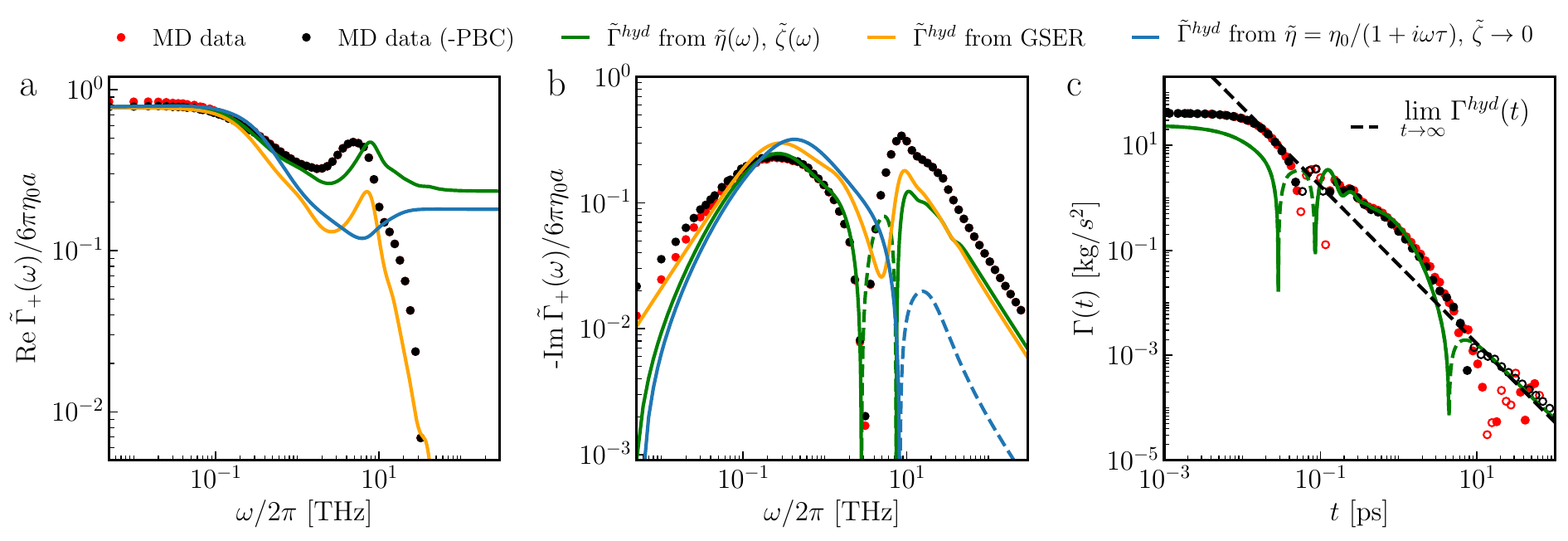}
		\caption{Comparison between the frequency-dependent friction of a SPC/E water molecule from the GLE, $\tilde{\Gamma}_+(\omega)$ (symbols), and the hydrodynamic prediction, $\tilde{\Gamma}^{hyd}(\omega)$ (lines). (a, b) Real and imaginary part of the Fourier transformed friction $\Tilde{\Gamma}_+(\omega)$ (circles,Appendix \ref{app:memkernel_iter}). The red and black circles denote the MD results without and with finite-size correction (Appendix \ref{app:pbc_correction}), respectively. We compute $\tilde{\Gamma}^{hyd}(\omega)$ according to Eq.~\eqref{eq:resp_sphere} (green) for $a = 141.36$ pm and $\hat{b} = b/a = 0.72$ and using the shear and volume viscosity parameters given in Appendix \ref{app:fit_params}. For the yellow lines we neglect compressibility, i.e., $c \to \infty$, and shear wave effects, i.e., $\alpha \to 0$, which corresponds to the GSER (Eq.~\eqref{eq:gser}). The blue lines show the hydrodynamic friction using a Maxwell model fit of the shear viscosity to the MD friction data and vanishing volume viscosity $\Tilde{\zeta} = 0$. The fitting constants are $\eta_{0}$ = 0.70 mPa s and $(2\pi\cdot\tau)^{-1 }$ = 0.43 THz. (c) Time domain behavior of the water memory kernel in (a, b). We compare the MD data (circles) with the numerical inverse Fourier transformation of the hydrodynamic friction $\Tilde{\Gamma}^{hyd}(\omega)$ in Eq.~\eqref{eq:resp_sphere} (green). The black dashed line is the predicted hydrodynamic tail in Eq.~\eqref{eq:hydro_tail_slip}. Dashed lines and empty circles denote negative values. The MD data for times higher than 10 ps is smoothed to reduce numerical noise.}
		\label{fig:resp_fct_kernel_water_comparison}
	\end{figure*}
	
	We insert the shear and volume viscosity spectra from the SPC/E water model MD simulations in Fig.~\ref{fig:visc_water} into Eq.~\eqref{eq:resp_sphere} to compute the hydrodynamic friction $\Tilde{\Gamma}^{hyd}(\omega)$  of a sphere. 
	To obtain a feeling for the influence of slip and sphere size, we
	in Fig.~\ref{fig:resp_fct_visc_freq} show $\Tilde{\Gamma}^{hyd}(\omega)$ for different values of the sphere radius $a$ and slip coefficient $\hat{b}=b/a$ (solid lines). 
	Note that the results are normalized by $6\pi a \eta_0$ with $\eta_0 = \sum_{j} \eta_{0,j} = 0.70$ mPa s, which is the steady-state friction for zero slip.
	For small radii $\sim a = 10^{-10}$ m, the friction $\Tilde{\Gamma}^{hyd}(\omega)$ in Fig.~\ref{fig:resp_fct_visc_freq} exhibits similar features as the frequency-dependent shear viscosity in Fig.~\ref{fig:visc_water} and in particular shows a peak around 7 THz, indicated by vertical dashed lines in Fig.~\ref{fig:resp_fct_visc_freq}. This behavior is abscent if a constant shear viscosity is assumed, as shown in Appendix \ref{app:rf_no_freq}.
	The slip length $b$ has a rather mild effect on the low-frequency friction, as it mostly modulates the absolute value.
	As $\omega \to 0$, the real part goes to $6\pi\eta_0 a$ for $\hat{b} \to 0$ and to $4\pi\eta_0 a$ for $\hat{b} \to \infty$ \cite{stokes1851effect}, as indicated by horizontal red and green bars to the left in Fig.~\ref{fig:resp_fct_visc_freq}. The real part of the friction functions does not decay to zero as $\omega \rightarrow \infty$ but instead reaches a plateau, the value of which depends on whether the slip parameter is zero or not. For radii $a > 10^{-10}$, the imaginary part interestingly changes its sign from negative to positive values. 
	In Appendix \ref{app:asymptotics}, we discuss the asymptotic behavior of the friction function for low and high frequencies. We observe that the imaginary part switches its sign again, which is not visible in the frequency range shown in Fig.~\ref{fig:resp_fct_visc_freq}.\\
	\indent We compare the results with different approximations of the full hydrodynamic theory. Short-dashed lines in Fig.~\ref{fig:resp_fct_visc_freq} denote the limiting scenario of infinity sound velocity $c\to \infty$, which represents the friction in an incompressible fluid. We observe distinct deviations from the full theory in the real and imaginary parts, which increase with increasing frequency. The high-frequency scaling depends strongly on whether we have a finite or a vanishing slip coefficient. In the incompressible case, the imaginary part diverges with $\simeq  \rho_0 a^2 \omega/(9\eta_0)$, corresponding to the so-called added mass term in the force acting on the sphere (see Appendix \ref{app:hydro_tail} for details). This term has been a long-standing issue in literature, since it causes a discontinuity in the initial value of the velocity autocorrelation function $C^{vv}(t)$ from the equipartition theorem, i.e., $C^{vv}(0)=k_BT/m$, to $k_BT/(m + m_0)$, where $m_0 = \frac{2}{3}\pi\rho_0a^3$ is the added mass of the half of the displaced fluid. The added mass term vanishes in the full hydrodynamic theory with compressibility, in agreement with previous works \cite{zwanzig1970hydrodynamic,chow1973brownian,mizuta2019bridging,chakraborty2011velocity}.\\ 
	\indent Dotted lines in Fig.~\ref{fig:resp_fct_visc_freq} denote the incompressible case with a constant shear viscosity, i.e., $\Tilde{\eta} = \eta_0$. The friction, in this case, significantly differs from the case of frequency-dependent shear viscosity, in particular the distinct peak around 7 THz is absent. 
	The long-dashed lines represent the case $c\to \infty$ and $\alpha \to 0$, where the hydrodynamic friction converges to the so-called generalized Stokes-Einstein relation (GSER)
	\begin{align}
		\label{eq:gser}
		\Tilde{\Gamma}^{hyd}(\omega) = 6\pi\Tilde{\eta}(\omega) a \frac{1+2\hat{b}}{1+3\hat{b}},
	\end{align}
	which is, in general with the additional approximation $b \to 0$, widely used in the context of rheological theory \cite{mason1995optical, squires2010fluid, schmidt2024nanoscopic}. As seen from Eq.~\eqref{eq:gser}, the GSER prediction is proportional to the shear viscosity spectra but differs significantly and from the full hydrodynamic theory, in particular for larger radii and in the high-frequency regime. \\
	\indent In Fig.~\ref{fig:resp_fct_visc_freq_volume_visc}, we analyze the dependence of the hydrodynamic friction on the volume viscosity $\Tilde{\zeta}(\omega)$. We compare the hydrodynamic friction for a finite frequency-independent volume viscosity $\zeta_0 = \sum_{j} \zeta_{0,j} = 1.69$ mPa s (dashed lines), 
	for vanishing volume viscosity $\zeta_0 = 0$, corresponding to the Stokes hypothesis (dotted lines) 
	and for the frequency-dependent volume viscosity extracted from MD simulations in Fig.~\ref{fig:visc_water} (solid lines).
	For large frequencies, we see that the real part of the friction diverges for constant volume viscosity but goes to a constant when using the full frequency-dependent volume viscosity from MD simulations or a vanishing volume viscosity (Appendix \ref{app:asymptotics}). 
	This shows that compression effects dominate the friction at high frequencies if an erroneous constant volume viscosity is assumed. 
	Remarkably, the predicted friction function in the low-frequency regime changes only marginally if, instead of the full frequency-dependent volume viscosity, a vanishing volume viscosity is assumed, demonstrating that the Stokes assumption \cite{stokes1880theories}, i.e., the neglect of volume viscosities, 
	is a very accurate approximation. However, the deviation between both cases increases for higher frequencies.
	\subsection{Comparison of the hydrodynamic and GLE friction for water and the hydrodynamic tail}
	
	In the following, we compare the friction 
	$\Gamma(t)$, defined  by the GLE in Eq.~\eqref{eq:GLE} and extracted from the MD simulations for SPC/E water (Appendix \ref{app:memkernel_iter}) with the hydrodynamic prediction.
	In Fig.~\ref{fig:resp_fct_kernel_water_comparison}(a, b) we 
	compare the real and imaginary parts of $\Tilde{\Gamma}_+(\omega)$ from  the GLE (circles) with predictions from the hydrodynamic Eq.~\eqref{eq:resp_sphere} 
	using different approximations: the green line shows $\Tilde{\Gamma}^{hyd}(\omega)$ using the frequency-dependent shear and volume viscosities extracted from MD simulations, 
	the yellow line shows $\Tilde{\Gamma}^{hyd}(\omega)$ using the frequency-dependent shear 
	viscosity $\Tilde{\eta}(\omega) $ but neglecting compressibility, i.e., $c \to \infty$, and shear-wave effects, i.e., $\alpha \to 0$, which corresponds to the GSER in Eq.~\eqref{eq:gser}, 
	the blue lines show the hydrodynamic friction using a Maxwell model for the shear viscosity, i.e., $\Tilde{\eta}(\omega) = \eta_0/(1+i\omega\tau)$ and vanishing volume viscosity $\Tilde{\zeta} = 0$.
	We compare the bare MD result for the memory kernel (red circles) with the result obtained by subtracting the effect of periodic boundary conditions (PBC) \cite{scalfi_pbc} (black circles, Appendix \ref{app:pbc_correction}). Note that here we show results from an extended MD simulation of SPC/E water compared to the results shown in Fig.~\ref{fig:visc_water} (Appendix \ref{app:MD_sim}), for the purpose of reducing statistical noise in the long-time behavior.
	\begin{figure*}
		\centering
		\includegraphics[width=1
		\linewidth]{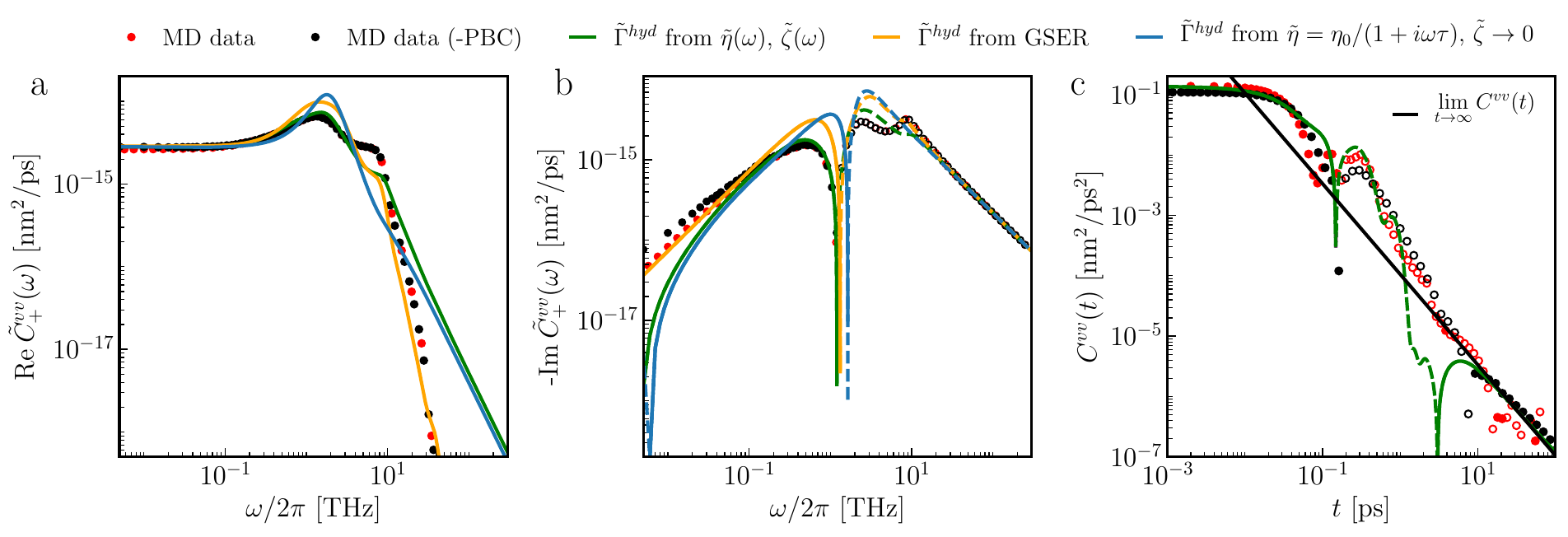}
		\caption{Comparison between the frequency-dependent velocity autocorrelation function $\Tilde{C}^{vv}_+(\omega)$ of a water molecule in water from MD simulations (symbols) with the hydrodynamic prediction according to Eq.~\eqref{eq:c_vv_kernel_ft}, using the results $\Tilde{\Gamma}_+(\omega)$ from hydrodynamic theory shown in Fig.~\ref{fig:resp_fct_kernel_water_comparison}. The color coding is the same as in Fig.~\ref{fig:resp_fct_kernel_water_comparison}. The black line in (c) is the predicted hydrodynamic tail in Eq.~\eqref{eq:hydro_tail_corrv}.}
		\label{fig:corrv_water_comparison}
	\end{figure*}
	\\ \indent The friction $\Tilde{\Gamma}_+(\omega)$ exhibits the same features as the hydrodynamic prediction $\Tilde{\Gamma}^{hyd}(\omega)$ from Eq.~\eqref{eq:resp_sphere} (green). Especially a peak around 6-8 THz is visible in both with a slight shift in frequency space. Here, we use $a = 141.36$ pm and $\hat{b} = 0.72$, which we obtain from fitting the MD results in the frequency domain, as described in Appendix \ref{app:slip_fit}.
	The prediction using the full frequency-dependent shear and volume viscosities (green) agrees rather nicely with the friction directly extracted from the GLE, though. Deviations between the MD data and the green lines become noticeable for frequencies above 10 THz, where the hydrodynamic prediction converges to a plateau in the real part and the MD data decay to zero.  These deviations are not unexpected \cite{zwanzig1970hydrodynamic} since standard hydrodynamic theory can not correctly describe
	the local interaction of the sphere
	with its environment for high frequencies which are mediated by smooth intermolecular
	forces \cite{straube2020rapid, seyler_molecular_2023} and that are not described accurately by an abrupt boundary condition at the spherical surface.
	Thus, we find that the friction can be predicted very well, including complex shear and volume viscosity models, below a frequency of 10 THz. The shear viscosity spectrum gives rise to the resonant feature in the friction kernel $\Tilde{\Gamma}_+(\omega)$ around 7 THz, which is missed by the Maxwell model used in previous models (blue line) \cite{zwanzig1970hydrodynamic, metiu1977hydrodynamic}. 
	\\ \indent The residual discrepencies below 10 THz could be caused by a modified viscosity near the particle surface since hydration shells may contribute to deviations of the macroscopic shear viscosity near the molecule. In Appendix \ref{app:lj_fluid}, we show that the agreement between hydrodynamic theory and MD is even better for a simpler system such as a Lennard-Jones fluid.\\
	\indent Interestingly, the good agreement for the low-frequency behavior in the friction function assumes non-negligible slip $b \neq 0$. In Appendix \ref{app:slip_fit}, we discuss the influence of the radius $a$ and the slip length $b$ on the hydrodynamic friction.
	The fitted values we obtain for the sphere radius $a = 141.36$ pm and the slip length $b  = 102.1$ pm (Appendix \ref{app:slip_fit}) are in a realistic range around 1 $\mathrm{\AA}$. 
	We obtain a water molecule mass of $m \approx 3\cdot10^{-26}$ kg from the equipartition theorem, i.e., $m C^{vv}(0) = k_BT$. For a density of $\rho_0 \approx 10^3$ kg/m$^{3}$, the estimate of the hydrodynamic radius accounting to $m = \frac{4}{3}\pi\rho_0 a^3$ is $a \approx$ 192.76 pm, in rough agreement with the estimated radius. The estimated radius and slip length are comparable with results of $a \approx 0.15$ nm and $b \approx 0.10$ nm obtained from determining the translational and rotational diffusion constants for SPC/E water \cite{zendehroud_slip}.
	\begin{figure*}
		\centering
		\includegraphics[width=1
		\linewidth]{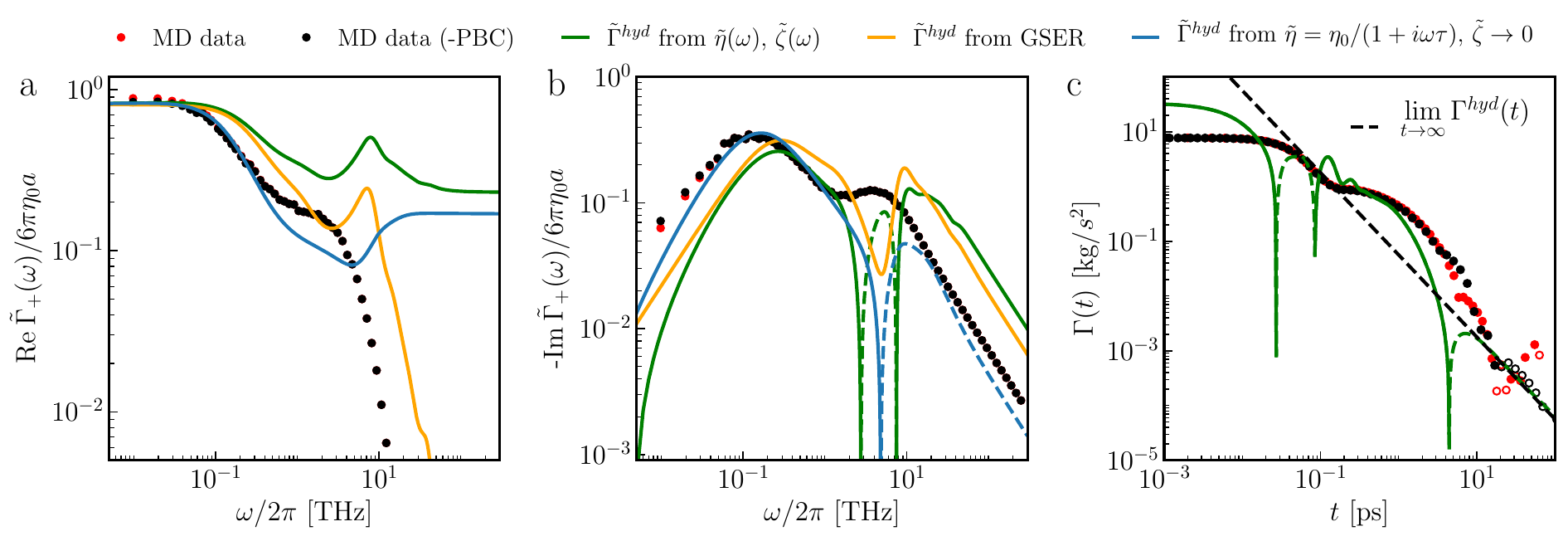}
		\caption{Comparison between the frequency-dependent friction of a methane molecule in water extraction from MD simulations using the GLE in Eq.~\eqref{eq:GLE} and the hydrodynamic prediction in Eq.~\eqref{eq:resp_sphere}. (a, b) Real and imaginary part of the Fourier transformed friction $\Tilde{\Gamma}_+(\omega)$ (time resolution 2 fs, data from \cite{kowalik2019memory}). The red and black circles denote the MD results without and with finite-size correction (Appendix \ref{app:pbc_correction}), respectively. We compare the computed friction with the results from Eq.~\eqref{eq:resp_sphere} (green) for $a = 139.29$ pm and $\hat{b} = b/a = 0.45$ and the shear and volume viscosity parameters given in Appendix \ref{app:fit_params}. The yellow line corresponds to the GSER (Eq.~\eqref{eq:gser}). The blue lines show the hydrodynamic friction using a Maxwell model fit of the shear viscosity to the MD friction data and vanishing volume viscosity $\Tilde{\zeta} = 0$. The fitting constants are $\eta_{0}$ = 0.70 mPa s and $(2\pi\cdot\tau)^{-1 }$ = 0.17 THz. (c) Time domain behavior of the methane memory kernel in (a, b). We compare the MD data (circles) with the numerical inverse Fourier transformation of the full hydrodynamic friction $\Tilde{\Gamma}^{hyd}(\omega)$ in Eq.~\eqref{eq:resp_sphere}. The black dashed line is the predicted hydrodynamic tail in Eq.~\eqref{eq:hydro_tail_slip}.}
		\label{fig:resp_fct_kernel_methane_comparison}
		
	\end{figure*}
	\\ \indent In an unbounded fluid, hydrodynamic backflow effects lead for long times to a power-law decay of the memory kernel as $\lim \limits_{t \to \infty} \Gamma^{hyd}(t) \approx - 3a^2\sqrt{\pi\eta_0\rho_0}t^{-3/2}$ \cite{ernst1970asymptotic, corngold1972behavior,chow1972effect}, which follows from our expression of the hydrodynamic friction in Eq.~\eqref{eq:resp_sphere} by assuming negligible slip, i.e., $b\rightarrow 0$, and vanishing compressibility, i.e., $c \rightarrow \infty$ (Appendix \ref{app:hydro_tail}). However, previous MD simulations of solute particles in liquid environments \cite{daldrop2017external,straube2020rapid} did not observe the predicted tail, rather showing a positive sign or a different power law behavior. Here we use hydrodynamic theory and finite-size effects to explain the absence of the tail, the influence of compressibility, slip, and frequency-dependent viscosity on the hydrodynamic tail has already been discussed in \cite{felderhof2005backtracking}. Since, as we show in Appendix \ref{app:hydro_tail_compress}, compressibility does not influence the hydrodynamic tail but is only relevant on intermediate time scales, we assume $\lambda \rightarrow 0$. As shown in Appendix \ref{app:asymptotics}, for low frequencies the hydrodynamic memory kernel in Eq.~\eqref{eq:resp_sphere} can be expanded as
	\begin{align}
		\label{eq:hydro_kernel_low_freq} 
		\frac{\Tilde{\Gamma}^{hyd}(\omega)}{6\pi\eta_0a} \simeq & \frac{1+2\hat{b}}{1+3\hat{b}} + \frac{a(1+2\hat{b})^2\sqrt{i\omega \rho_0/\eta_0}} {\sqrt{2}(1+3\hat{b})^2} + \mathcal{O}(\omega^{3/2}).
	\end{align}
	We see that the $\omega^{1/2}$-term, which is responsible for the $t^{-3/2}$ power-law decay, is rescaled by the slip length $\hat{b}$. Higher-order terms are influenced by the slip length $\hat{b}$ and the model we choose for the shear viscosity $\Tilde{\eta}(\omega)$. 
	By inverse Fourier transformation, one obtains for long times
	\begin{equation}
		\label{eq:hydro_tail_slip} 
		\lim \limits_{t \to \infty} \Gamma^{hyd}(t) \approx - 3a^2 \frac{(1+2\hat{b})^2}{(1+3\hat{b})^2}\sqrt{\pi\eta_0\rho_0}t^{-3/2},
	\end{equation}
	shown as a black dashed line in Fig.~\ref{fig:resp_fct_kernel_water_comparison}(c).
	The long-time behavior of the water memory kernel is governed by a competition between the hydrodynamic tail and the long-time behavior of the frequency-dependent shear viscosity we use. The power-law tail in the MD data only agrees with Eq.~\eqref{eq:hydro_tail_slip} if we subtract the masking finite-size effects (black circles). The uncorrected MD data (red circles) are dominated by PBC effects and do not exhibit the long-time tail in Eq.~\eqref{eq:hydro_tail_slip} at time scales readable with MD simulations \cite{scalfi_pbc}. 
	
	\subsection{Long-time tail of velocity autocorrelation function}
	\begin{figure*}
		\centering
		\includegraphics[width=.8
		\linewidth]{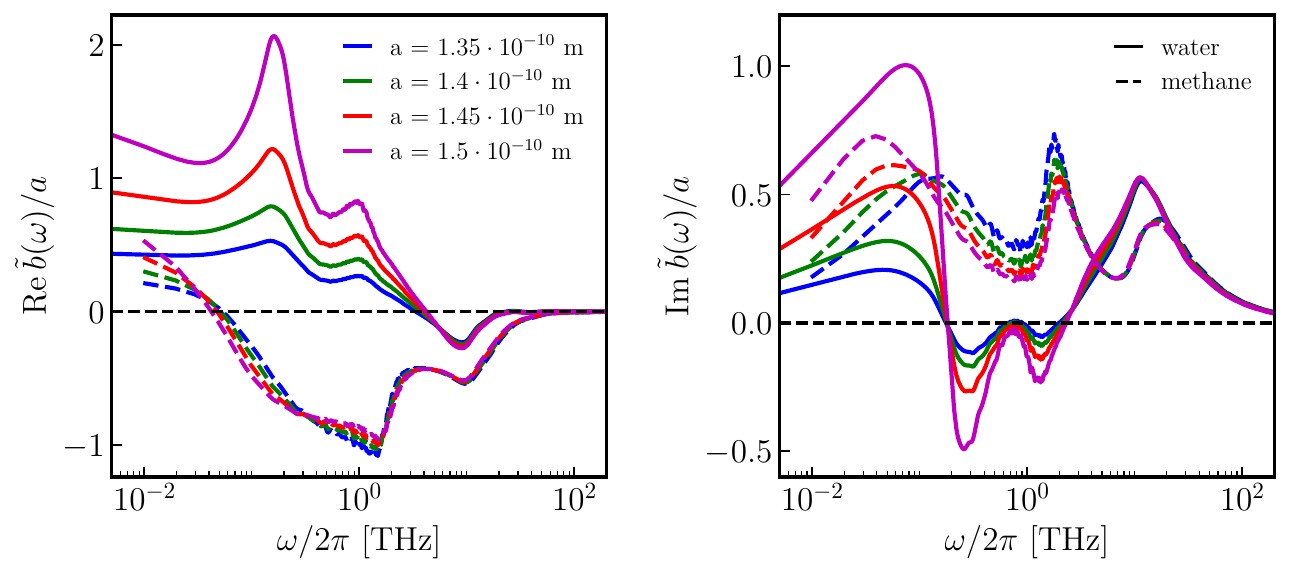}
		\caption{Frequency-dependent slip length $\Tilde{b}(\omega) = \:\text{Re}\:\Tilde{b}(\omega) + i\:\text{Im}\:\Tilde{b}(\omega)$ obtained by assuming the frequency-dependent friction kernel $\Tilde{\Gamma}_+(\omega)$ extracted from MD data shown in Fig.~\ref{fig:resp_fct_kernel_water_comparison} and \ref{fig:resp_fct_kernel_methane_comparison} to be equivalent to the hydrodynamic expression in Eq.~\eqref{eq:resp_sphere} and solving the equation for the slip length $\Tilde{b}(\omega)$. We show calculated data for different radii $a$. Solid lines denote the results for the water memory kernel and dashed lines for the methane memory kernel.}
		\label{fig:resp_fct_kernel_slip_profiles}
		
	\end{figure*}
	The velocity autocorrelation function $C^{vv}(t)$ (VACF) of a solute particle was used in various studies to compare hydrodynamic and stochastic theories \cite{zwanzig1970hydrodynamic,metiu1977hydrodynamic,chakraborty2011velocity,mizuta2019bridging}.
	The GLE in Eq.~\eqref{eq:GLE} relates the memory kernel and the VACF by a Volterra equation (Appendix \ref{app:memkernel_iter}), which can be written as (derived in Appendix \ref{app:deriv_cvv_ft})
	\begin{equation}
		\label{eq:c_vv_kernel_ft}
		\Tilde{C}_+^{vv}(\omega) = \frac{i \omega k_BT}{i \omega \Tilde{\Gamma}_+(\omega) - m \omega^2}.
	\end{equation}
	In Fig.~\ref{fig:corrv_water_comparison}, we compare predictions according to Eq.~\eqref{eq:c_vv_kernel_ft} using hydrodynamic theory with MD results for the VACF. As for the memory kernel, we achieve good agreement in the frequency domain up to frequencies of 10 THz with the full theory in Eq.~\eqref{eq:resp_sphere} (green line). The hydrodynamic tail of the VACF for long times (Appendix \ref{app:hydro_tail})
	\begin{equation}
		\label{eq:hydro_tail_corrv} 
		\lim \limits_{t \to \infty} C^{vv}(t) \approx  \frac{k_BT \sqrt{\rho_0}}{12} (\pi\eta_0 t )^{-3/2},
	\end{equation}
	agrees with the prediction from hydrodynamic theory and the MD data with finite-size correction (black circles) for long times $> 10$ ps in Fig.~\ref{fig:corrv_water_comparison}(c). These findings in the VACF support that the best prediction of the solute particle dynamics is achieved by including the macroscopic viscosity spectra of the fluid into the hydrodynamic theory.\\
	\subsection{Results for a methane molecule moving in water}
	\label{app:methane}
	In Fig.~\ref{fig:resp_fct_kernel_methane_comparison}, we show the memory kernel $\Tilde{\Gamma}_+(\omega)$ for a methane molecule in SPC/E water extracted from MD simulations, again with and without PBC and finite-size correction. The data is taken from the work of Kowalik \textit{et al.} \cite{kowalik2019memory}. The methane molecule in these simulations
	is modeled as a monoatomic Lennard-Jones sphere, which makes it possible to compute the friction using Eq.~\eqref{eq:resp_sphere} for a spherical particle.
	In Fig.~\ref{fig:resp_fct_kernel_methane_comparison}(a, b), we see that the real and imaginary parts of the Fourier transformed friction $\Tilde{\Gamma}_+(\omega)$ 
	do not show the same features as the prediction from the hydrodynamic equations (green and yellow). In particular, the pronounced peak in the real part at around 7-8 THz arising from the shear viscosity of water is absent, and the memory kernel from MD in the time domain in Fig.~\ref{fig:resp_fct_kernel_methane_comparison}(c) does not contain any noticeable oscillations, in contrast to the hydrodynamic prediction.
	\\\indent Thus, we observe considerable discrepancies between the friction extracted directly from MD simulations and the hydrodynamic prediction for a hydrophobic molecule in a water environment. Constant slip effects on the methane surface cannot explain these differences. Rather, the local shear viscosity near the methane molecule seems to differ from the bulk shear viscosity. This is in line with a recent analysis of tracer-particle dynamics
	in hydrogels, where the assumption of a thin interfacial shell with viscoelastic properties different from bulk would reconcile the experimental measurements with hydrodynamic predictions \cite{schmidt2024nanoscopic}.

	\subsection{Frequency-dependent surface slip from MD simulations}
	The observation that the hydrodynamic theory does not accurately predict the friction kernel extracted directly from the MD data above 10 THz in Fig.~\ref{fig:resp_fct_kernel_water_comparison} and \ref{fig:resp_fct_kernel_methane_comparison}, and that a good agreement is rather achieved with the simplified GSER model, is a clear sign that the hydrodynamic theory 
	used by us misses key aspects of molecular friction. A reasonable extension is to assume the slip coefficient $b$, which follows from the Navier boundary condition at the spherical surface (Appendix \ref{app:friction_deriv}) to be frequency-dependent as well, i.e., $b = \Tilde{b}(\omega)$. Note that a frequency-dependent surface friction coefficient, i.e., $\Tilde{\Lambda}(\omega) = \Tilde{\eta}(\omega)/\Tilde{b}(\omega)$, has been assumed in many works \cite{bocquet1993hydrodynamic,bocquet1993hydrodynamic,omori2019full,oga2021theoretical}. To obtain the slip spectrum, we assume the frequency-dependent memory kernel $\Tilde{\Gamma}_+(\omega)$ to equal the hydrodynamic expression in Eq.~\eqref{eq:resp_sphere} and solve for the slip length $\Tilde{b}(\omega)$. The results we show in Fig.~\ref{fig:resp_fct_kernel_slip_profiles} for different radii $a$ are, therefore, the slip spectra that lead to a perfect agreement between the friction from the MD data we show in Fig.~\ref{fig:resp_fct_kernel_water_comparison} and \ref{fig:resp_fct_kernel_methane_comparison} and the hydrodynamic prediction, and exhibit interesting features. For example, we observe regions where the real part of the slip coefficient is negative, which would correspond to a negative real part of the surface friction coefficient $\Tilde{\Lambda}(\omega)$, contrary to standard linear response theory. Negative slip coefficients indicate a local shear viscosity around the spherical particle different from the bulk viscosity \cite{uematsu2018analytical}.\\
	\indent For the water particle's friction, denoted as solid lines, a distinct positive peak in the real part around 0.2 THz is visible, which is in the region of the shear viscosity mode I of water (Appendix \ref{app:fit_params}) and relates to an induced slip effect due to hydrogen network topology changes. The real part of the slip length of a water molecule at the prominent peak of shear viscosity mode III around 7-8 THz is negative, suggesting that hydrogen stretching modes cause local viscosities around the tracer particle that are higher than the macroscopic shear viscosity. For methane, denoted by dashed lines in Fig.~\ref{fig:resp_fct_kernel_slip_profiles}, the real part is negative in nearly the whole frequency range. This agrees with the fact that methane in water is surrounded by a highly-ordered solvation and presumably high-viscosity shell structure \cite{daldrop2017external}. 
	
	\section{Conclusions}
	We demonstrate that predictions from macroscopic hydrodynamic theory are in good agreement with the friction coefficient
	directly extracted via the GLE from MD simulations of a single water molecule in liquid water if the frequency dependence of the shear and volume viscosities is properly accounted for. 
	This establishes the long-sought link between macroscopic fluid hydrodynamics and the molecular friction in a fluid. 
	We also show that it is important to include the frequency-dependent volume viscosity of the fluid that
	has the asymptotic behavior  $\Tilde{\zeta} \rightarrow 0$ as $\omega \rightarrow \infty$. \\
	\indent Interestingly, the agreement between the friction from the molecular water motion (obtained via the GLE)  and the hydrodynamic prediction is achieved without including spatial or wave-vector dependencies 
	of the viscosity functions.
	Nevertheless, we cannot exclude the possibility that such spatial dependencies are present, 
	especially at high frequencies.
	We have mostly dealt with the homogeneous case, where the moving molecule is identical 
	to the surrounding fluid molecules.
	In contrast, we observe pronounced discrepancies between the friction obtained from hydrodynamic theory and simulations for the
	inhomogeneous case of a methane molecule moving in water, especially at frequencies above 10 THz. 
	By calculating the frequency-dependent surface slip from the MD simulations, we conclude that our current  hydrodynamic model neglects the modified viscous properties 
	of the water solvation layer around a methane molecule. 
	It would therefore be desirable to develop inhomogeneous hydrodynamic models for the friction of host molecules in liquids in the presence
	of solvation shells that exhibit viscosity properties that are different from the bulk.
	\section*{Acknowledgements}
	This work was supported by the Deutsche Forschungsgemeinschaft (DFG) via Grant No. SFB 1449 'Dynamic Hydrogels at Biointerfaces', Project A02,
	and Grant No. SFB 1114 'Scaling Cascades in Complex Systems', Project C02. The
	authors would like to acknowledge the HPC Service of ZEDAT, Freie Universität Berlin, for providing computing time.
	
	%\section*{Appendix}
	\appendix
	
	\section{Derivation of the Friction of a Sphere from the Transient Stokes Equation}
	\label{app:friction_deriv}
	The Navier-Stokes equation reads
	\begin{eqnarray}
		\label{eq:mom_cons}
		\frac{\partial \rho (\Vec{r},t) v_i (\Vec{r},t)}{\partial t}  + \nabla_j\:\rho(\Vec{r},t) &&v_i(\Vec{r},t) v_j(\Vec{r},t)\\ \nonumber  = F_i(\Vec{r},t) &&+ \nabla_j \sigma_{ij}(\Vec{r},t),
	\end{eqnarray}
	where $i,j \in \{x,y,z\}$ and doubly appearing indices are summed over. The mass density $\rho$, velocity $v_i$ and  volume force $F_i$ are functions of time $t$ and position $\Vec{r}$.
	The symmetric stress tensor $\sigma_{ij}(\Vec{r},t)$ consists of a diagonal pressure contribution 
	and components that depend on velocity gradients, i.e., $\nabla_j v_i(\Vec{r},t)$ \cite{hansen1990theory,daivis1994comparison,j2007statistical,schulz2020molecular}. For a linear, homogeneous, isotropic compressible fluid, it is on the linear level given by 
	\begin{eqnarray}
		\label{eq:stress_tensor}
		\sigma_{ij}(\Vec{r},t)  \:= &&
		- P(\Vec{r},t)\delta_{ij}  \\ \nonumber  &&+ \int \int \Bigl[ \eta(|\Vec{r}'|,t')\Bigl(\nabla_i v_j(\Vec{r}- \Vec{r}',t-t') \\ \nonumber
		&&+ \nabla_j v_i(\Vec{r}- \Vec{r}',t-t')\Bigr)
		\\\nonumber + \delta_{ij}\Bigl(\zeta(|\Vec{r}'|,t') && - \frac{2}{3}\eta(|\Vec{r}'|,t')\Bigr) \nabla_k v_k(\Vec{r}- \Vec{r}',t-t')\bigr]\:d\Vec{r}' dt',
	\end{eqnarray}
	where $P$ is the pressure and $\eta$ and $\zeta$ are the shear and volume viscosity kernels, which in general are
	time- and space-dependent. If the viscosity kernels decay on length- and time scales that are small compared to those on which $\nabla_j v_i(\Vec{r},t)$ varies, one can approximate the stress tensor in Eq.~\eqref{eq:stress_tensor} as
	\begin{eqnarray}
		\nonumber
		\sigma_{ij}&& (\Vec{r},t) \approx  -  P(\Vec{r},t) \delta_{ij} +
		\left( \zeta_0- \frac{2}{3}\eta_0\right) \delta_{ij} \nabla_k v_k(\Vec{r},t) \\ \label{eq:stress_tensor_nofreq} &&+ \eta_0   \Bigl(\nabla_i v_j(\Vec{r},t) + \nabla_j v_i(\Vec{r},t)             
		\Bigr),
	\end{eqnarray}
	where $\eta_0$ and $\zeta_0$ are the time- and space-integrated viscosities
	\begin{align}
		\eta_0 =  \int \int \eta(|\Vec{r}'|,t') d\Vec{r}' dt',\\
		\zeta_0 =  \int \int \zeta(|\Vec{r}'|,t') d\Vec{r}' dt'.
	\end{align}
	We define a fluid with a stress tensor given by Eq.~\eqref{eq:stress_tensor_nofreq} as a Newtonian fluid, i.e., the viscosities are time- and space-dependent and the stress tensor is linear in the velocity gradients, 
	but will explicitly consider viscoelastic fluids in this work, i.e. the viscosity depends on the history of the velocity gradients (Eq.~\eqref{eq:stress_tensor}).
	If we neglect the non-linear term in the Navier-Stokes equation (second term on the left-hand side in Eq.~\eqref{eq:mom_cons}), which is justified for low Reynolds numbers, and use the expression of the stress tensor in Eq.~\eqref{eq:stress_tensor}, we arrive at the linear transient Stokes equation
	\begin{eqnarray}
		\label{eq:Stokes}
		\rho(\Vec{r},t) \frac{\partial v_i(\Vec{r},t)}{\partial t} &&= F_i(\Vec{r},t) - \nabla_i P(\Vec{r},t)  \\ \nonumber
		+\int \int\Bigl(\frac{1}{3}\eta(|\Vec{r}'|,t') &&+ \zeta(|\Vec{r}'|,t')\Bigr) \nabla_i\nabla_k v_k(\Vec{r}- \Vec{r}',t-t')d\Vec{r}' dt'\:\\ \nonumber + \int \int\eta(|\Vec{r}'|,t') &&\nabla_k \nabla_k v_i(\Vec{r}- \Vec{r}',t-t')d\Vec{r}' dt'.
	\end{eqnarray}
	The frequency-dependent friction of a sphere can be calculated using the Green's function
	of the Stokes equation in Eq.~\eqref{eq:Stokes} \cite{kim2013microhydrodynamics,erbacs2010viscous}. 
	To derive the Green's function, we take the divergence of Eq.~\eqref{eq:Stokes} and obtain
	\begin{eqnarray}
		\label{eq:Stokes2} 
		\nonumber
		\nabla_i^2 P(\Vec{r},t) - \frac{\partial^2 P(\Vec{r},t)}{c^2 \partial t^2 } &&= \nabla_i F_i(\Vec{r},t) +\int \int\Bigl(\frac{4}{3}\eta(|\Vec{r}'|,t') \\  + \zeta(|\Vec{r}'|,t')\Bigr) \nabla_i^2 && \nabla_k v_k(\Vec{r} - \Vec{r}',t-t')d\Vec{r}' dt',
	\end{eqnarray}
	where we used the linearized continuity equation, i.e., $\rho_0 \nabla_i (\partial v_i/\partial t) = - \partial^2 \rho /\partial t^2$, 
	and the linearized isentropic equation of state  $\rho - \rho_0 = c^{-2}(P-P_0)$, where the speed of sound is denoted by $c$, 
	from which  follows that $\partial^2 \rho /\partial t^2 = c^{-2} \partial^2 P /\partial t^2$.
	Fourier-transforming Eqs.~\eqref{eq:Stokes} and~\eqref{eq:Stokes2} we obtain
	\begin{eqnarray}
		\label{eq:Stokes_FT}
		i\omega\rho_0\Tilde{v}_i(\Vec{k},\omega) && = \Tilde{F}_i(\Vec{k},\omega) - i k_i \Tilde{P}(\Vec{k},\omega) \\ \nonumber  - [\Tilde{\eta}(\Vec{k},\omega)/3 &&+ \Tilde{\zeta}(\Vec{k},\omega)]k_ik_j\Tilde{v}_j(\Vec{k},\omega) - \Tilde{\eta}(\Vec{k},\omega)k_jk_j\Tilde{v}_i(\Vec{k},\omega),
	\end{eqnarray}
	and 
	\begin{eqnarray}
		\label{eq:Stokes_FT2}
		(\frac{\omega^2}{c^2} - k_ik_i)\Tilde{P}(\Vec{k},\omega) &&= i k_i\Tilde{F}_i(\Vec{k},\omega) \\ \nonumber - i[4 \Tilde{\eta}(\Vec{k},\omega)/3 &&+ \Tilde{\zeta}(\Vec{k},\omega)]k_ik_ik_j\Tilde{v}_j(\Vec{k},\omega).
	\end{eqnarray}
	Note that the viscosity kernels $\eta$ and $\zeta$ are both single-sided in the time domain, i.e., $\eta(\Vec{r},t) = 0$ and $\zeta(\Vec{r},t) = 0$ for $t < 0$. 
	We next assume that both viscosities decay quickly in space, 
	so that their Fourier transforms become  independent of $k$, 
	i.e., $\Tilde{\eta}(\Vec{k},\omega) \rightarrow  \Tilde{\eta}(\omega)$ and $\Tilde{\zeta}(\Vec{k},\omega) \rightarrow  \Tilde{\zeta}(\omega)$. 
	Solving  Eq.~\eqref{eq:Stokes_FT2} for $\Tilde{P}$ and inserting into Eq.~\eqref{eq:Stokes_FT}, we arrive at an equation for the velocity as a function of the external force \cite{erbacs2010viscous}. To solve this equation, we decompose the velocity into the transverse and  longitudinal parts, 
	i.e., $\Tilde{v}_i(\Vec{k},\omega) = \Tilde{v}_i^T(\Vec{k},\omega) + \Tilde{v}_i^L(\Vec{k},\omega)$, which fulfill $k_i \Tilde{v}_i^T(\Vec{k},\omega) = 0$ and $k_i\Tilde{v}_i(\Vec{k},\omega) = k_i \Tilde{v}_i^L(\Vec{k},\omega)$.
	In Fourier space, the Green's function $\Tilde{G}_{ij}$ of the velocity is defined by
	\begin{equation}
		\Tilde{v_i}(\Vec{k},\omega) = \Tilde{G}_{ij}(\Vec{k},\omega) \Tilde{F}_j(\Vec{k},\omega),
	\end{equation}
	and is a sum of transverse and longitudinal components, i.e., $\Tilde{G}_{ij}(\Vec{k},\omega) = \Tilde{G}_{ij}^T(\Vec{k},\omega) + \Tilde{G}_{ij}^L(\Vec{k},\omega)$. The transverse part describes the velocity field in the incompressible case and accounts for shear effects. It is given by
	\begin{equation}
		\label{eq:greens_trans}
		\Tilde{G}_{ij}^T(\Vec{k},\omega) = \frac{(\delta_{ij} - k_ik_j/k^2)/\Tilde{\eta}(\omega)}{k^2 + \alpha^2(\omega)},
	\end{equation}
	where the length scale $\alpha^{-1}$ is defined as
	\begin{equation}
		\alpha^2(\omega) = i \omega\rho_0/\Tilde{\eta}(\omega).
	\end{equation}
	The longitudinal component describes compression effects and reads
	\begin{equation}
		\label{eq:greens_long}
		\Tilde{G}_{ij}^L(\Vec{k},\omega) = \frac{k_ik_j\lambda^2(\omega)}{\Tilde{\eta}(\omega) \alpha^2(\omega)k^2(k^2+\lambda^2(\omega))}.
	\end{equation}
	The length scale $\lambda^{-1}$ is defined as
	\begin{equation}
		\lambda^2(\omega) = \frac{i \omega\rho_0}{4\Tilde{\eta}(\omega)/3 + \Tilde{\zeta}(\omega) - i\rho_0c^2/\omega}.
	\end{equation}
	\\The full frequency-dependent Green's function $G_{ij}(\Vec{r},\omega) = G_{ij}^T(\Vec{r},\omega) + G_{ij}^L(\Vec{r},\omega)$ in real space reads
	\begin{eqnarray}
		\label{eq:Greensfct_app}
		G_{ij}(\Vec{r},\omega) &&= \frac{1}{4\pi\Tilde{\eta}\alpha^2r^3}\{\delta_{ij}([1+r\alpha+r^2\alpha^2]e^{-r\alpha} \\ \nonumber &&- [1+r\lambda]e^{-r\lambda})
		-3\hat{r}_i\hat{r}_j([1+r\alpha+r^2\alpha^2/3]e^{-r\alpha} \\ \nonumber
		&&- [1+r\lambda+r^2\lambda^2/3]e^{-r\lambda})\}. 
	\end{eqnarray}
	The asymptotic behavior of the Green's function, which is discussed in detail in \cite{erbacs2010viscous}, strongly depends on the inverse length scales $\alpha = \alpha(\omega)$ and $\lambda = \lambda(\omega)$. Note that in \cite{erbacs2010viscous} a different definition of the Fourier transformation is used and frequency-independent viscosities are assumed. 
	\\To calculate the friction acting on an oscillating sphere, we have to compute the Green's function for the stress tensor, denoted by $\sigma_{ijk}$, defined as
	\begin{equation}
		\sigma_{ij}(\Vec{r},\omega) = \sigma_{ijk}(\Vec{r},\omega) F_k(\Vec{r},\omega).
	\end{equation}
	Without derivation and referring to \cite{erbacs2010viscous}, the stress tensor Green's function is given by
	\begin{eqnarray}
		\label{eq:stress}
		\sigma_{ijk}(\Vec{r},\omega)/\Tilde{\eta}(\omega) &&= G_{ijk}(\Vec{r},\omega) + G_{jik}(\Vec{r},\omega) \\  \nonumber &&+ (\alpha^2/\lambda^2 - 2)G_{llk}(\Vec{r},\omega)\delta_{ij},
	\end{eqnarray}
	where $\nabla_k G_{ij} = G_{kij}$. To obtain the fluid velocity around a sphere with radius $a$, we use a standard singularity Ansatz \cite{kim2013microhydrodynamics}
	\begin{equation}
		\label{eq:hydro_force}
		G_{ij}^{sp}(\Vec{r},\omega) = (C_0 + C_2 a^2 \nabla_k \nabla_k) G_{ij}(\Vec{r},\omega),
	\end{equation}
	where the velocity field around the sphere follows as $\Tilde{v}_i^{sp}(\Vec{r},\omega) = \Tilde{F}_j(\omega)G_{ij}^{sp}(\Vec{r},\omega)$, with $\Tilde{F}_j$ being a force source. 
	We choose the coefficients $C_0$ and $C_2$ such that the boundary conditions on the spherical surface are satisfied. If we assume a finite slip at the spherical surface, we can split the boundary conditions at the surface into a kinematic and a Navier boundary condition. The kinematic boundary condition at $|r| = a$ can be written as
	\begin{equation}
		\label{eq:kinematic_bc}
		6\pi\Tilde{\eta}(\omega)a\hat{r}_iG^{sp}_{ij}(\omega) = \hat{r}_j,
	\end{equation}
	which defines the sphere velocity $\Tilde{V}_i^{sp}(\omega) = \Tilde{F}_i(\omega)/6\pi\Tilde{\eta}(\omega)a$. Note that only in the zero-frequency limit the source force $\Tilde{F}_i(\omega)$ equals the actual force on the sphere. The Navier boundary condition for the tangential stress at $|r| = a$ reads
	\begin{align} \nonumber
		b[\nabla_k G_{ij}^{sp}(\omega) &+ \nabla_i G_{kj}^{sp}(\omega)]\hat{r}_k\mathcal{L}_{li} \\ \label{eq:navier_bc}
		&= [G_{ij}^{sp}(\omega) - \delta_{ij}/6\pi\Tilde{\eta}(\omega)a]\mathcal{L}_{li},
	\end{align} 
	where $b$ is the slip length and we define the projection operator as $\mathcal{L}_{li} = (\delta_{li} - \hat{r}_l\hat{r}_i)$.
	
	The final result for $G_{ij}^{sp}(\Vec{r},\omega)$ reads, using Eq.~\eqref{eq:Greensfct_app}
	\begin{align}
		\label{eq:sphere_greens_app}
		G_{ij}^{sp}(\Vec{r},\omega) = & \frac{1}{4\pi\Tilde{\eta}(\omega)\alpha^2 r^3}\\ \nonumber 
		\cdot\:\Bigl\{\delta_{ij}&(E_1[1+r\alpha+r^2\alpha^2]e^{-r\alpha} - E_2[1+r\lambda]e^{-r\lambda}) \\ \nonumber
		- & 3\hat{r}_i\hat{r}_j(E_1[1+r\alpha+r^2\alpha^2/3]e^{-r\alpha}  \\ \nonumber
		-  &E_2[1+r\lambda+r^2\lambda^2/3]e^{-r\lambda}) \Bigr\},
	\end{align}
	with the coefficients
	\begin{align}
		\label{eq:e1}
		E_1 &= \frac{2}{3}e^{\hat{\alpha}}\frac{(1+2\hat{b})(3+3\hat{\lambda}+\hat{\lambda}^2)}{W}, \\
		\label{eq:e2}
		E_2 &= \frac{2}{3}e^{\hat{\lambda}}\frac{(1+2\hat{b})(3+3\hat{\alpha}+\hat{\alpha}^2) + \hat{b}\hat{\alpha}^2(1+\hat{\alpha})}{W},
	\end{align}
	and 
	\begin{equation}
		W = (2+2\hat{\lambda}+\hat{\lambda}^2)(1+\hat{b}(3+\hat{\alpha}))+(1+\hat{\alpha})(1+2\hat{b})\hat{\lambda}^2/\hat{\alpha}^2.
	\end{equation}
	We define the dimensionless slip length, $\hat{b} = b/a$, and the dimensionless decay constants $\hat{\alpha} = a\alpha$ and $\hat{\lambda} = a\lambda$. The corresponding friction function $\Tilde{\Gamma}^{hyd}(\omega)$ is given by
	\begin{equation}
		\delta_{ij}\Tilde{\Gamma}^{hyd}(\omega) = \frac{\Tilde{F}_i^{sp}(\omega)}{\Tilde{V}_j^{sp}(\omega)} = -6\pi\Tilde{\eta}(\omega) a\int d^3 r \hat{r}_k \sigma_{kij}\delta(|r|-a),
	\end{equation}
	where $\Tilde{V}_j^{\text{sp}}$ is the frequency-dependent velocity amplitude of the sphere. For the hydrodynamic force $\Tilde{F}_{i}^{sp}(\omega)$ on a spherical particle, we use the projection of the stress tensor on the surface and integrate it over the sphere surface.
	Using Eq.~\eqref{eq:stress}, and the derivative of $G_{ij}^{sp}(\Vec{r},\omega)$ in Eq.~\eqref{eq:sphere_greens_app}, we obtain the friction function of the spherical particle
	\begin{eqnarray}
		\label{eq:resp_sphere_app} \nonumber
		\Tilde{\Gamma}^{hyd}(\omega) = &&  \frac{4\pi\Tilde{\eta}(\omega)a}{3} W^{-1}\Bigl\{(1+\hat{\lambda})(9+9\hat{\alpha}+\hat{\alpha}^2)(1+2\hat{b}) \\ && + (1+\hat{\alpha})[2\hat{\lambda}^2(1+2\hat{b})+\hat{b}\hat{\alpha}^2(1+\hat{\lambda})]\Bigr\},
	\end{eqnarray}
	where we use the identities $\int d^3r \delta_{ij} \delta(|r|-a) = 4\pi a^2 \delta_{ij}$ and $\int d^3r \hat{r}_i \hat{r}_j \delta(|r|-a) = 4\pi a^2 \delta_{ij}/3$.
	Assuming negligible slip, i.e., $b\rightarrow 0$, and vanishing compressibility, i.e., $\lambda \rightarrow 0$, we have
	\begin{align}
		\label{eq:resp_sphere_app2}
		\Tilde{\Gamma}^{hyd}(\omega) = 6\pi a \Tilde{\eta}(\omega) (1+a\alpha+a^2\alpha^2/9).
	\end{align}
	Thus, the frequency-dependent hydrodynamic force on a spherical
	particle, i.e., $\Tilde{F}_{i}^{sp}(\omega) = - \Tilde{\Gamma}^{hyd}(\omega) \Tilde{V}_i^{sp}(\omega)$, is in the same limit given by
	\begin{align}
		\label{eq:mom_sphere12}
		\Tilde{F}_{i}^{sp}(\omega) = -6\pi a \Tilde{\eta}(\omega) \Tilde{V}_i^{sp}(\omega) (1+a\alpha+a^2\alpha^2/9).
	\end{align}
	
	\section{Fluid Momentum around a Moving Sphere}
	\label{app:momentum_sphere}
	From the velocity field, i.e., $\Tilde{v}_i^{sp}(\Vec{r},\omega) = \Tilde{F}_j(\Vec{r},\omega)G_{ij}^{sp}(\Vec{r},\omega)$, and the expression in Eq.~\eqref{eq:sphere_greens_app}, we can calculate the fluid momentum outside the moving sphere as
	\begin{align}
		\label{eq:mom_sphere1}
		\Tilde{p}_{i}(\omega) = \rho_0 \int_{|r|>a} \Tilde{v}_i^{sp}(\Vec{r},\omega) d^3r.
	\end{align}
	We assume that the force source is oscillating along the x-direction, i.e., $\Vec{\Tilde{F}}(\omega) = (\Tilde{F}(\omega),0,0)^T$, so that the momentum points in the x-direction. The volume integral in Eq.~\eqref{eq:mom_sphere1} involves the angular integrals
	\begin{align}
		\label{eq:volume_int}
		\int_{|r|>a} d^3r \delta_{ij} &= \int_a^\infty r^2 dr \int_0^\pi d\theta\:\text{sin} \theta \int_0^{2\pi} d \Phi\:\delta_{ij}, \\ \nonumber &= 4 \pi\delta_{ij} \int_a^\infty r^2 dr, \\ \nonumber
		\int_{|r|>a} d^3r \hat{r}_i \hat{r}_j  &= \int_a^\infty r^2 dr \int_0^\pi d\theta\:\text{sin}\theta\:\text{cos}^2 \theta \int_0^{2\pi} d \Phi \:\delta_{ij}, \\ & \nonumber= 2\pi \delta_{ij} \int_a^\infty r^2 dr \int_{-1}^1 du\:u^2, \\
		&= \frac{4}{3} \pi \delta_{ij} \int_a^\infty r^2 dr.
	\end{align}
	In the derivation of the flow field around a sphere in Appendix \ref{app:friction_deriv}, we use the kinematic boundary condition, i.e., $6\pi a \Tilde{\eta}\hat{r}_i G_{ij}^{sp} = \hat{r}_j$ for $|r| = a$ , in Eqs.~\eqref{eq:kinematic_bc} and \eqref{eq:navier_bc} and define the sphere velocity $\Tilde{V}_i^{sp}(\omega) = \Tilde{F}_i(\omega)/6\pi\Tilde{\eta}(\omega)a$, such that in the low-frequency (steady) limit and without slip, the source force equals the actual force on the sphere. Using this definition, the identities in Eq.~\eqref{eq:volume_int} and the expression in Eq.~\eqref{eq:sphere_greens_app} and inserting them into Eq.~\eqref{eq:mom_sphere1}, we arrive at
	\begin{align}
		\label{eq:mom_sphere2}
		\Tilde{p}_{i}(\omega) &= 6\pi\frac{\rho_0}{\alpha^2} a \Tilde{V}_i^{sp}(\omega) \int_a^\infty \frac{1}{r} dr \\ \nonumber
		\cdot\:\Bigl\{\delta_{ij}&(E_1[1+r\alpha+r^2\alpha^2]e^{-r\alpha} - E_2[1+r\lambda]e^{-r\lambda}) \\ \nonumber
		- & \delta_{ij}(E_1[1+r\alpha+r^2\alpha^2/3]e^{-r\alpha}  \\ \nonumber
		-  &E_2[1+r\lambda+r^2\lambda^2/3]e^{-r\lambda}) \Bigr\},
	\end{align}
	\begin{align}
		\label{eq:mom_sphere3}
		& = 6\pi\frac{\rho_0}{\alpha^2} a \Tilde{V}_i^{sp}(\omega) \int_a^\infty dr \Bigl[ \frac{2}{3} E_1 r \alpha^2 e^{-r\alpha} +  \frac{1}{3} E_2 r \lambda^2 e^{-r\lambda} \Bigr],
	\end{align}
	\begin{align}
		\label{eq:mom_sphere4}
		= - 6\pi\frac{\rho_0}{\alpha^2} a \Tilde{V}_i^{sp}(\omega) \Bigl[\frac{2}{3} &E_1 [e^{-r\alpha}(r\alpha + 1)]|_a^\infty \\ \nonumber + \frac{1}{3} &E_2 [e^{-r\lambda}(r\lambda + 1)]|_a^\infty \Bigr],\\
		\label{eq:mom_sphere5}
		= 6\pi\frac{\rho_0}{\alpha^2} a \Tilde{V}_i^{sp}(\omega) \Bigl[\frac{2}{3}& E_1 [e^{-a\alpha}(a\alpha + 1)] \\ \nonumber + \frac{1}{3} &E_2 [e^{-a\lambda}(a\lambda + 1)] \Bigr],
	\end{align}
	
	For $b\rightarrow 0$ and $\lambda \rightarrow 0$, the constants $E_1$ and $E_2$ in Eqs.~\eqref{eq:e1} and \eqref{eq:e2} become $E_1 = e^{a\alpha}$ and $E_2 = (1+a\alpha + a^2\alpha^2/3)$, and we obtain for the momentum
	\begin{align}
		\label{eq:mom_sphere6}
		\Tilde{p}_{i}(\omega) = 6\pi\frac{\rho_0}{\alpha^2} a \Tilde{V}_i^{sp}(\omega) (1+a\alpha+a^2\alpha^2/9).
	\end{align}
	The force is given by $\Tilde{F}_{i} = i\omega \Tilde{p}_{i}$, which leads to
	\begin{align}
		\label{eq:mom_sphere6}
		\Tilde{F}_{i}(\omega) = 6\pi a \Tilde{\eta}(\omega) \Tilde{V}_i^{sp}(\omega) (1+a\alpha+a^2\alpha^2/9),
	\end{align}
	which is identical to the result obtained in Eq.~\eqref{eq:mom_sphere1} from integrating the surface force over the oscillating sphere. Thus, the net frequency-dependent momentum of the fluid inside the sphere has to vanish, and we have no added mass due to the motion of the liquid inside the sphere. It follows that the friction $\Tilde{\Gamma}^{hyd}$ calculated from hydrodynamic theory equals the friction $\Gamma(t)$ extracted from single-particle trajectories using the GLE, and no fluid mass correction has to be applied.
	
	\section{Simulation Setup}
	\label{app:MD_sim}
	
	We perform all MD simulations using the GROMACS simulation package \cite{pronk2013gromacs} (version 2020-Modified). For water, we use the SPC/E \cite{berendsen1987missing} and TIP4P/2005 \cite{Abascal_2005} rigid water models. We pre-equilibrate the system in an NPT ensemble ($P$ = 1 bar) using a Berendsen barostat \cite{Berendsen_1984} set to 1 atm. For production runs, we perform all simulations in the NVT ensemble with a temperature of $T$ = 300 K, controlled with a velocity rescaling thermostat \cite{bussi2007canonical}. For electrostatics, we use the particle-mesh Ewald method \cite{Darden_1993}, with a cut-off length of 1 nm. We allow simulations to run for 600 ns, using integration time steps of 1 fs. We perform simulations in a 3.5616 nm cubic box with 1250 water molecules. 
	We use the trajectories of two traced water particles for the memory kernel extraction. For the results we show in Appendix \ref{app:memkernel_time_res}, we additionally run simulations with integration time steps of 2 and 4 fs.\\
	\indent For the results we show in Fig.~\ref{fig:resp_fct_kernel_water_comparison}, we run an MD simulation of SPC/E water with a total length of 1 $\mu s$ and a time step of 2 fs, and use the trajectories of 15 water particles for the memory kernel extraction.\\
	\indent For the Lennard-Jones (LJ) fluid, we simulate a system at $T=92$ K with a box length of 5 nm and 2744 LJ particles. For the particles, we took the Lennard-Jones parameters
	of argon of the GROMOS53a6 force field \cite{Oostenbrink_2004} ($\sigma$ =
	3.410 $\mathrm{\AA}$, $\epsilon$ = 0.996 kJ/mol and a cutoff radius of 2.5$\sigma$). Using LJ units, the systems are at $T^*$ = 0.77 and $P^*$ = 0.04
	corresponding to the liquid phase \cite{vrabec2006comprehensive,ahmed2010effect}. The system is first equilibrated in the NPT ensemble ($P$ = 17 bar)
	followed by a production run in the NVT ensemble for
	10 ns with a time step of 2 fs. We use the trajectories of 50 LJ particles for the memory kernel extraction.
	\section{Calculation of Frequency-Dependent Shear and Volume Viscosity Spectra from MD Simulations}
	\label{app:green_kubo}
	The shear viscosity kernel $\eta(t)$ is given by the trace-free part of the stress tensor by the Green-Kubo relation \cite{hansen1990theory,j2007statistical,zwanzig1965time,schulz2020molecular}
	\begin{eqnarray}
		\label{eq:Kubo_shear_visc}
		\Tilde{\eta}(\Vec{k} = 0, \omega) =&& \int_0^{\infty}dt\:\eta(t)\:e^{-i\omega t}, \\ =&&\nonumber \frac{V}{6 k_BT}\int_0^{\infty}e^{-i\omega t} \sum_{i\neq j}\langle \Pi_{ij}(t) \Pi_{ij}(0)\rangle dt,
	\end{eqnarray}
	where $V$ is the volume of the fluid. We define the trace-free part of the stress tensor $\sigma_{ij}$ as
	\begin{equation}
		\Pi_{ij} = \sigma_{ij} - \delta_{ij}\frac{1}{3} \sum_{k} \sigma_{kk},
	\end{equation}
	where $i,j \in \{x,y,z\}$.
	For the computation of the shear viscosity spectrum, we use Eq.~\eqref{eq:Kubo_shear_visc} by calculating the time-correlation functions of the stress tensor entries and applying the half-sided Fourier transform.\\
	\indent Employing the Green-Kubo relations, we use the fluctuations of the instantaneous pressure from its average value $\langle{P}\rangle$, i.e., $\delta P(t) = P(t) - \langle P \rangle$, to compute the volume viscosity kernel $\zeta(t)$. $P(t)$ is computed from the trace of the stress tensor, i.e.,  $P(t) = \frac{1}{3}\sum_{k} \sigma_{kk}(t)$. Using the half-sided Fourier transformation, we compute the volume viscosity spectrum via \cite{medina2011molecular}
	\begin{eqnarray}
		\label{eq:Kubo_volume_visc}
		\Tilde{\zeta}(\Vec{k} = 0, \omega) =&& \int_0^{\infty}dt\:\zeta(t)\:e^{-i\omega t}, \\\nonumber =&& \frac{V}{k_BT}\int_0^{\infty}e^{-i\omega t} \langle \delta P(t) \delta P(0) \rangle dt.
	\end{eqnarray}
	For the Fourier transformation of the viscosity data, and the memory kernel data as well, we use the FFT algorithm implemented in NumPy v.~1.18.5 \cite{harris2020array}, where we assume the input signal $x(t)$ to be single-sided, i.e., $x(t<0)$ = 0. All data in the time domain are truncated at 10 ps. Note that the FFT in NumPy uses the opposite convention for the Fourier transformation definition we use here.
	
	\section{Fitting of the Viscosity Data}
	\label{app:fitting_procedure}

	We fit the real part of the Fourier-transformed viscosity spectra $\Tilde{\eta}(\omega)$ and $\Tilde{\zeta}(\omega)$ extracted from the MD simulation by a combination of six and five exponential-oscillating components according to Eqs.~\eqref{eq:visc_model} and \eqref{eq:bulk_model}, respectively, using the Levenberg-Marquardt\:algorithm implemented in scipy v.~1.4 \cite{2020SciPy-NMeth}. The initial values for all $\eta_{0,j}$, $\tau_{n,j}$ and $\tau_{o,j}$ are chosen suitably; the same for $\zeta_{0,j}$, $\tau_{v,j}$ and $\tau_{w,j}$. We constrain the parameter space to positive values. We filter the data set beforehand on a logarithmic frequency scale to reduce the overall data points to fit. We also weight the data exponentially so that the data for small frequencies become more important for the fit. After optimizing the parameters, we use them as initial parameters to peform the final fit of the data in the time domain. Here the input data are filtered on logarithmic time-domain scale but without exponential weighting. This allows us to fit the low- and high-frequency regimes very well at the same time. 
	The obtained fit parameters are summarized in Appendix \ref{app:fit_params}.
	
	\newpage
	\section{Summary of Fitting Parameters}
	\label{app:fit_params}
	
	%\newpage\thispagestyle{empty}
	\begin{table}[h]
		\centering
		\caption{Fitting parameters for the viscoelastic model of the shear viscosity in Eqs.~\eqref{eq:visc_model_time} and \eqref{eq:visc_model} to MD data of the SPC/E water model and the TIP4P/2005 water model. The time scales are converted to frequencies for ease of comparison with Fig.~\ref{fig:visc_water}(a, b, c).}
		\begin{ruledtabular}
			\begin{tabular}{c  c  c } 
				
				Parameter & SPC/E & TIP4P/2005 \\ 
				\hline
				$\eta_{0,I}$ & 0.45 mPa s& 0.25 mPa s\\
				$(2\pi\cdot\tau_{n,I})^{-1 }$ & 1.96 THz & 1.83 THz \\
				$(2\pi\cdot\tau_{o,I})^{-1 }$ & 0.67 THz & 0.47 THz \\
				\hline
				$\eta_{0,II}$ & 0.14 mPa s& 0.06 mPa s\\
				$(2\pi\cdot\tau_{n,II})^{-1 }$ & 3.33 THz & 1.94 THz \\
				$(2\pi\cdot\tau_{o,II})^{-1 }$ & 1.69 THz & 1.54 THz \\
				\hline
				$\eta_{0,III}$ & 0.08 mPa s& 0.06 mPa s\\
				$(2\pi\cdot\tau_{n,III})^{-1 }$ & 5.45 THz & 4.16 THz \\
				$(2\pi\cdot\tau_{o,III})^{-1 }$ & 8.08 THz & 7.74 THz \\
				\hline
				$\eta_{0,IV}$ & 0.02 mPa s& 0.43 mPa s\\
				$(2\pi\cdot\tau_{n,IV})^{-1 }$ & 10.58 THz & 2.04 THz \\
				$(2\pi\cdot\tau_{o,IV})^{-1 }$ & 15.50 THz & 0.76 THz \\
				\hline
				$\eta_{0,V}$ & 0.005 mPa s& 0.02 mPa s\\
				$(2\pi\cdot\tau_{n,V})^{-1 }$ & 27.47 THz & 20.95 THz \\
				$(2\pi\cdot\tau_{o,V})^{-1 }$ & 31.74 THz & 17.03 THz \\
				\hline
				$\eta_{0,VI}$ & 3.31 $\cdot 10^{-5}$ mPa s& 8.11 $\cdot 10^{-4}$ mPa s\\
				$(2\pi\cdot\tau_{n,VI})^{-1 }$ & 6.31 THz & 15.61 THz \\
				$(2\pi\cdot\tau_{o,VI})^{-1 }$ & 42.10 THz & 39.73 THz \\
				
			\end{tabular}
		\end{ruledtabular}
		\label{tab:fit_visc}
	\end{table}
	\begin{table}[!htbp]
		\centering
		\caption{Fitting parameters for the viscoelastic model of the volume viscosity in Eqs.~\eqref{eq:bulk_model_time} and \eqref{eq:bulk_model} to MD data of the SPC/E water model and the TIP4P/2005 water model. The time scales are converted to frequencies for ease of comparison with Fig.~\ref{fig:visc_water}(d, e, f).}
		\begin{ruledtabular}
			\begin{tabular}{c  c  c } 
				
				Parameter & SPC/E & TIP4P/2005 \\ 
				\hline
				$\zeta_{0,I}$ & 0.75 mPa s& 0.23 mPa s\\
				$(2\pi\cdot\tau_{v,I})^{-1 }$ & 1.11 THz & 0.13 THz \\
				$(2\pi\cdot\tau_{w,I})^{-1 }$ & 0.41 THz & 0.09 THz \\
				\hline
				$\zeta_{0,II}$ & 0.45 mPa s& 1.23 mPa s\\
				$(2\pi\cdot\tau_{v,II})^{-1 }$ & 1.53 THz & 1.14 THz \\
				$(2\pi\cdot\tau_{w,II})^{-1 }$ & 0.75 THz & 0.50 THz \\
				\hline
				$\zeta_{0,III}$ & 0.04 mPa s& 0.01 mPa s\\
				$(2\pi\cdot\tau_{v,III})^{-1 }$ & 6.89 THz & 4.33 THz \\
				$(2\pi\cdot\tau_{w,III})^{-1 }$ & 7.07 THz & 7.10 THz \\
				\hline
				$\zeta_{0,IV}$ & 0.44 mPa s& 0.55 mPa s\\
				$(2\pi\cdot\tau_{v,IV})^{-1 }$ & 5.52 THz & 7.96 THz \\
				$(2\pi\cdot\tau_{w,IV})^{-1 }$ & 3.55 THz & 4.14 THz \\
				\hline
				$\zeta_{0,V}$ & 0.02 mPa s& 0.02 mPa s\\
				$(2\pi\cdot\tau_{v,V})^{-1 }$ & 20.76 THz & 19.39 THz \\
				$(2\pi\cdot\tau_{w,V})^{-1 }$ & 18.20 THz & 17.88 THz \\
			\end{tabular}
		\end{ruledtabular}
		\label{tab:fit_bulk_visc}
	\end{table}
	\newpage
	\newpage\thispagestyle{empty}
	\begin{table}
		\centering
		\caption{Fitting parameters for the viscoelastic model of the shear viscosity in Eqs.~\eqref{eq:visc_model_time},~\eqref{eq:visc_model} and volume viscosity in Eqs.~\eqref{eq:bulk_model_time},~\eqref{eq:bulk_model} to MD data of the LJ fluid. The time scales are converted to frequencies for ease of comparison with Fig.~\ref{fig:lj_results}(a, b).}
		\begin{ruledtabular}
			\begin{tabular}{c  c  c } 
				
				Parameter & $\eta$ & $\zeta$ \\ 
				\hline
				$\eta_{0,I}$, $\zeta_{0,I}$ & 0.12 mPa s& 0.23 mPa s\\
				$(2\pi\cdot\tau_{n,I})^{-1 }$, $(2\pi\cdot\tau_{v,I})^{-1 }$ & 2.38 THz & 70.71 THz \\
				$(2\pi\cdot\tau_{o,I})^{-1 }$,$(2\pi\cdot\tau_{w,I})^{-1 }$ & 1.18 THz & 40.69 THz \\
				\hline
				$\eta_{0,II}$, $\zeta_{0,II}$ & 0.05 mPa s& 0.05 mPa s\\
				$(2\pi\cdot\tau_{n,II})^{-1 }$, $(2\pi\cdot\tau_{v,II})^{-1 }$ & 4.04 THz & 6.12 THz \\
				$(2\pi\cdot\tau_{o,II})^{-1 }$, $(2\pi\cdot\tau_{w,II})^{-1 }$ & 2.41 THz & 3.06 THz \\
				\hline
				$\eta_{0,III}$, $\zeta_{0,III}$ & 0.07 mPa s& 0.32 mPa s\\
				$(2\pi\cdot\tau_{n,III})^{-1 }$, $(2\pi\cdot\tau_{v,III})^{-1 }$ & 32.87 THz & 15.32 THz \\
				$(2\pi\cdot\tau_{o,III})^{-1 }$, $(2\pi\cdot\tau_{w,III})^{-1 }$ & 19.25 THz & 11.09 THz \\
				\hline
				$\eta_{0,IV}$, $\zeta_{0,IV}$ & 0.01 mPa s& 0.01 mPa s\\
				$(2\pi\cdot\tau_{n,IV})^{-1 }$, $(2\pi\cdot\tau_{v,IV})^{-1 }$ & 449.15 THz & 987.59 THz \\
				$(2\pi\cdot\tau_{o,IV})^{-1 }$, $(2\pi\cdot\tau_{w,IV})^{-1 }$ & 333.42 THz & 18.07 THz \\
				
			\end{tabular}
		\end{ruledtabular}
		\label{tab:fit_visc_lj}
	\end{table}
	
	\section{Components of the Shear Viscosity for SPC/E Water}
	\label{app:fit_params_plot_spce}
	
	\begin{figure*}
		\centering
		\includegraphics[width=1
		\linewidth]{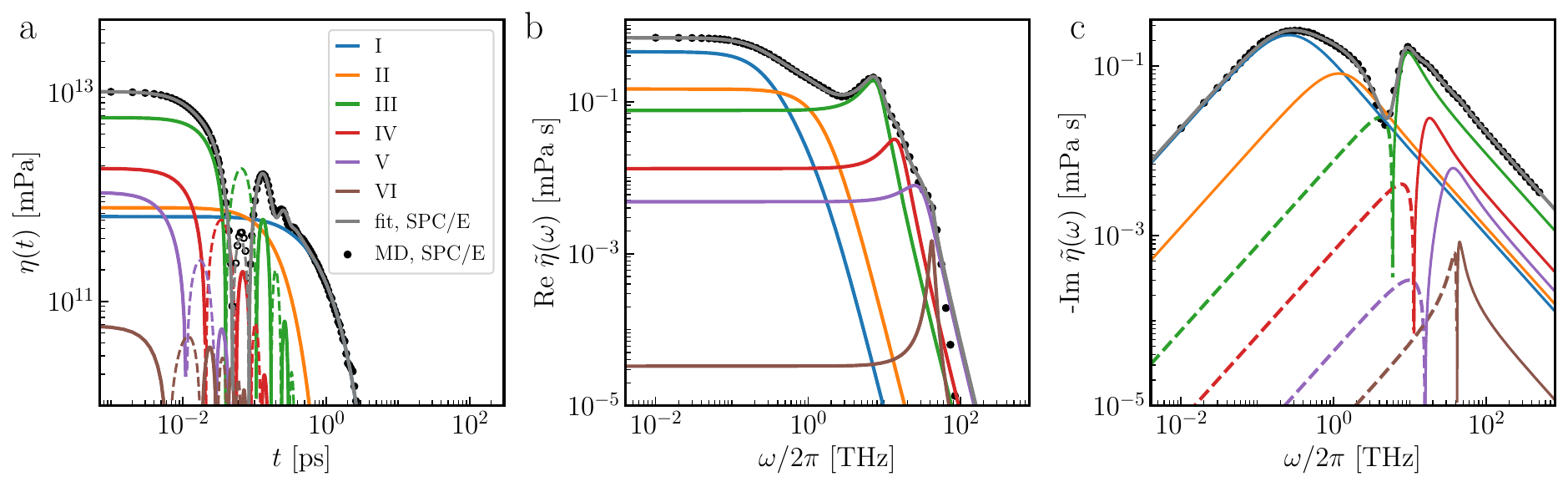}
		\caption{Extracted shear viscosity $\eta$ from MD simulations of SPC/E water (from Fig.~\ref{fig:visc_water}) together with the fitting components (lines) according to Eq.~\eqref{eq:visc_model_time} and  the fitting parameters in Table~\ref{tab:fit_visc} (Appendix \ref{app:fit_params}). Dashed lines and empty circles denote negative values.}
		\label{fig:visc_water_spce_comps}
	\end{figure*}

	\begin{figure*}
		\centering
		\includegraphics[width=1
		\linewidth]{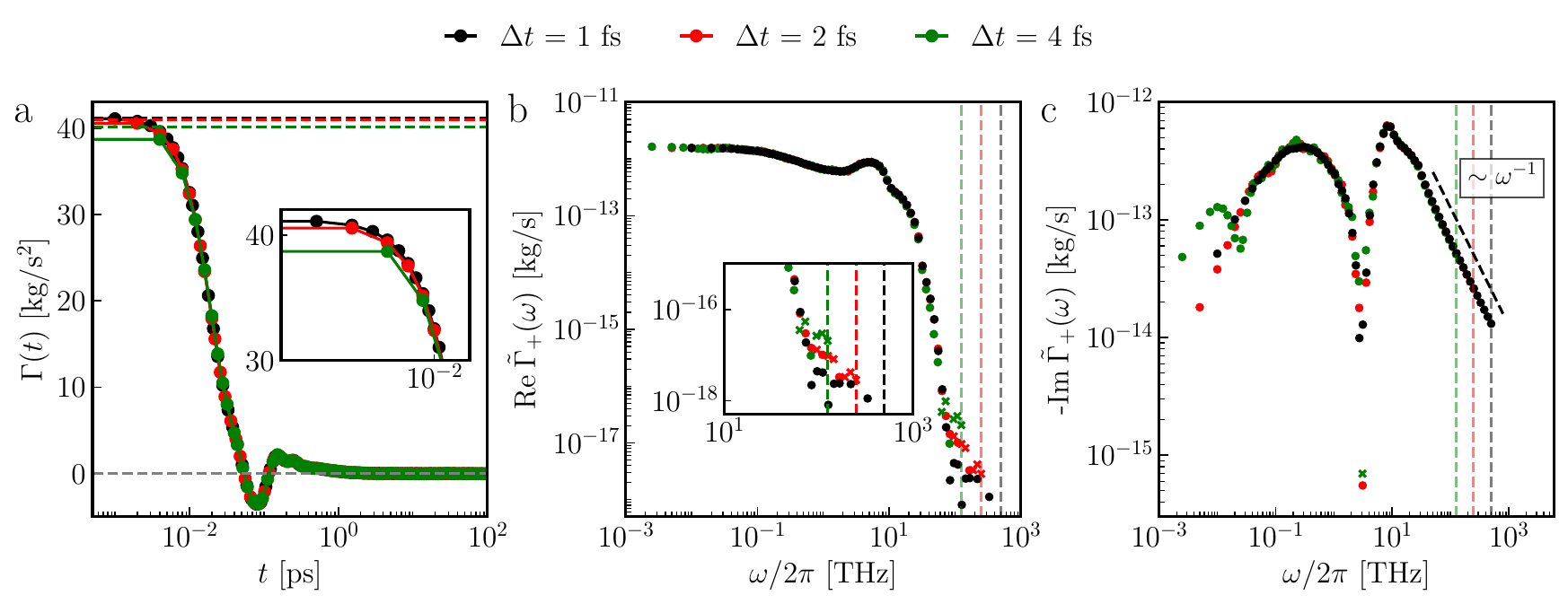}
		\caption{Influence of the time resolution on the frequency-dependent friction. We show the extracted water memory kernel $\Gamma(t)$ (a, extracted as explained in Appendix \ref{app:memkernel_iter}) and the real and imaginary part of the Fourier transformation $\Tilde{\Gamma}_+(\omega)$ (b, c) for SPC/E water MD simulations  with different time resolutions as explained in Appendix \ref{app:MD_sim}. In the frequency domain, circles denote positive values, and crosses denote negative values. The colored dashed lines in the time domain (a) denote the initial values of the memory kernel data $\Gamma(0)$. The colored dashed lines in the frequency domain in (b, c) denote the maximal frequencies of the Fourier-transformed data.}
		\label{fig:mem_kernel_time_res_spce}
	\end{figure*}

	The fitting components according to Eq.~\eqref{eq:visc_model_time} and the fitting parameters in Table~\ref{tab:fit_visc} (Appendix \ref{app:fit_params}) for the SPC/E water MD simulation are depicted in Fig.~\ref{fig:visc_water_spce_comps}, together with the MD data and the total fit shown in Fig.~\ref{fig:visc_water}.

	\section{Friction from MD Simulations with Different Time Resolution}
	\label{app:memkernel_time_res}
	
	In Fig.~\ref{fig:mem_kernel_time_res_spce}, we investigate the influence of the time resolution on the MD simulation, where we simulated SPC/E water for different time resolutions. The memory kernel, extracted as explained in Appendix \ref{app:memkernel_iter}, exhibits, besides numerical noise, no distinct differences between the different time resolutions. All important features we observe in the memory kernel for 1 fs are also visible for different time resolutions in Fig.~\ref{fig:mem_kernel_time_res_spce}. The same applies for the extracted viscosity spectra in Fig.~\ref{fig:visc_water}, since they exhibit similar molecular features as discussed in the main text.
	The deviating behavior from the exponential-oscillatory fitting model starting around 30-60 THz seen for the viscosity spectra occurs at lower resolutions in Fig.~\ref{fig:mem_kernel_time_res_spce}(b), which rules out discretization problems in this frequency range. At higher frequencies, in the resolution limit regime, the real part data in Fig.~\ref{fig:mem_kernel_time_res_spce}(b) is dominated by noise, which means that we cannot make a statement about the actual high-frequency scaling. For the 1 fs data (black), the data points become unstable around 100 THz, which is a fifth of the maximal frequency of 500 THz. 
	This suggests that our used FFT algorithm explained in Appendix \ref{app:fitting_procedure} is numerically unstable for the non-periodic data sets in this frequency regime, and data points in this regime should not be used for interpretation.
	
	\section{Friction of a Sphere for Frequency-Independent Viscosities}
	\label{app:rf_no_freq}
	\begin{figure*}
		\centering
		\includegraphics[width=1
		\linewidth]{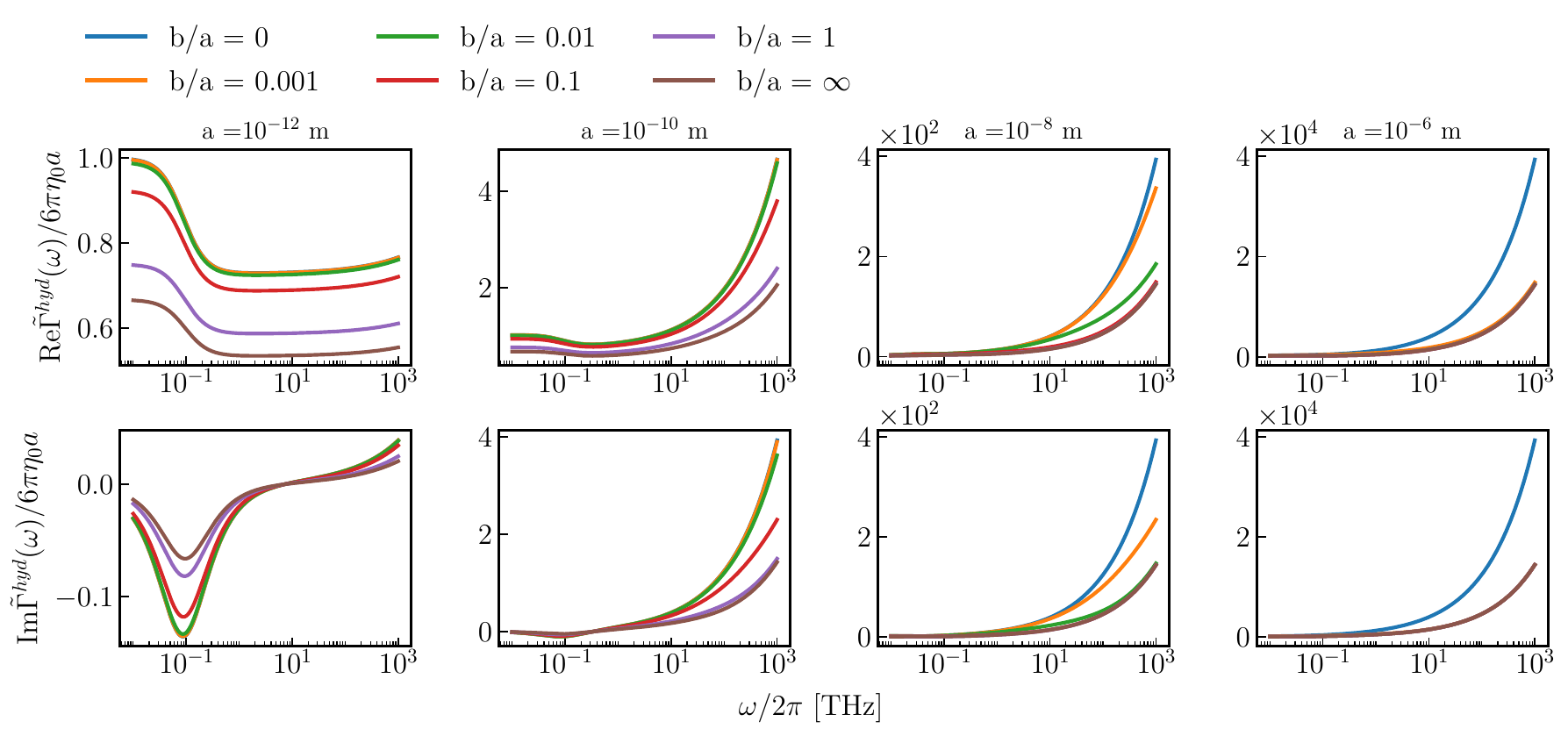}
		\caption{Real and imaginary parts of the rescaled friction function of a sphere $\Tilde{\Gamma}^{hyd}(\omega) = \:\text{Re}\:\Tilde{\Gamma}^{hyd}(\omega) + i\:\text{Im}\:\Tilde{\Gamma}^{hyd}(\omega)$ given by Eq.~\eqref{eq:resp_sphere}, for various slip parameters $\hat{b}$ and sphere radii $a$. Here we use a constant shear viscosity of $\eta_0 = 1 $ mPa s, the water density $\rho_0$ = 10$^3$ kg/m$^3$, $c=10^3\:$m/s, and vanishing volume viscosity $\Tilde{\zeta} = 0$ for comparability with the results in \cite{erbacs2010viscous}.}
		\label{fig:resp_fct_no_freq}
	\end{figure*}
	
	In Fig.~\ref{fig:resp_fct_no_freq}, we show the calculated hydrodynamic friction of a sphere (Eq.~\eqref{eq:resp_sphere}) for constant shear viscosity $\Tilde{\eta}(\omega) = \eta_0$ for different sphere radii $a$. Note that we use a vanishing volume viscosity $\Tilde{\zeta} = 0$, for better comparability with the results in \cite{erbacs2010viscous}. As already discussed in \cite{erbacs2010viscous}, the friction sensitively depends on the slip length $b$ and its real and imaginary parts both increase as $\omega \rightarrow \infty$. Note that the high-frequency behavior of the imaginary part differs from the results in \cite{erbacs2010viscous}, as we use a different definition of the Fourier transformation. We refer to \cite{erbacs2010viscous} for a detailed discussion of these results, but note that the addition of frequency-dependent shear and volume viscosity significantly changes the friction as shown in Fig.~\ref{fig:resp_fct_visc_freq}.
	
	\section{Low- and High-Frequency Scaling of the Hydrodynamic Friction of a Sphere}
	\label{app:asymptotics}
	In the following, we analyze  friction in Eq.~\eqref{eq:resp_sphere}, without considering the limiting cases $c\rightarrow\infty$, $\Tilde{\eta} = \eta_0$ or $\alpha \rightarrow 0$.
	For the real part of the friction in Eq.~\eqref{eq:resp_sphere}, we analytically obtain the asymptotic behavior (for finite $c$ and $b\neq0$), for shear viscosity and volume viscosity given by the models in Eqs.~\eqref{eq:visc_model} and \eqref{eq:bulk_model}, as
	\begin{align}
		\label{eq:rf_real_asympt}
		\frac{\:\text{Re}\:\Tilde{\Gamma}^{hyd}(\omega)}{6\pi\eta_0a} \simeq
		\begin{cases} 
			\omega \rightarrow 0 & \frac{1+2\hat{b}}{1+3\hat{b}} + \frac{a(1+2\hat{b})^2\sqrt{ \rho_0/\eta_0}}{\sqrt{2} (1+3\hat{b})^2} \omega^{1/2} \\ & + \mathcal{O}(\omega^{3/2}), \\
			\omega \rightarrow \infty & \\
			\text{if } \Tilde{\zeta} = \zeta_{0} & \frac{2a}{9}\Phi \frac{(\alpha_\infty)^2}{2\lambda_c}  \omega^{1/2},
			\\\text{otherwise} &  \frac{2a}{9} \Phi  \frac{(\alpha_\infty)^2}{\lambda_\infty} \omega^{0},
		\end{cases}
	\end{align}
	with the constants 
	\begin{align}
		\Phi &= (\displaystyle\sum_{j = I}^{VI} \frac{\eta_{0,j}\tau_{n,j}}{\tau_{o,j}^2})/\eta_{0}, \\ \eta_0 &= \sum_{j=I}^{VI} \eta_{0,j}.
	\end{align}
	
	Here, we introduced high-frequency limiting values for the inverse length scales $\alpha_\infty$, $\lambda_\infty$ and $\lambda_c$. They are determined via
	\begin{align}
		\alpha(\omega \to \infty) &\equiv \alpha_\infty = \sqrt{\rho_0\Tilde{\eta}_\infty} , \\
		\lambda_\infty &= \sqrt{\rho_0\Tilde{Z}_\infty}, \\
		\lambda_c &= \sqrt{\rho_0/2\displaystyle\sum_{j = I}^{V} \zeta_{0,j}}, \nonumber
	\end{align}
	where $\Tilde{\eta}_\infty$ and $\Tilde{Z}_\infty$ follow from the high-frequency limits of the shear and volume viscosities in Eqs.~\eqref{eq:visc_model} and \eqref{eq:bulk_model}, with $\Tilde{Z}(\omega) = 4\Tilde{\eta}(\omega)/3 + \Tilde{\zeta}(\omega) - i \rho_0 c^2/\omega $
	\begin{align}
		\Tilde{\eta}_\infty &= 1/ \left(\displaystyle\sum_{j = I}^{VI} \frac{\eta_{0,j}\tau_{n,j}}{\tau_{o,j}^2}\right), \\
		\Tilde{Z}_\infty &= 1/\left(\displaystyle\sum_{j = I}^{VI} \frac{4\eta_{0,j}\tau_{n,j}}{3\tau_{o,j}^2} + \displaystyle\sum_{j = I}^{V} \frac{\zeta_{0,j}\tau_{v,j}}{\tau_{w,j}^2} + \rho_0c^2\right).
	\end{align}
	For $\Tilde{\zeta}(\omega)\neq \Tilde{\zeta}_0$, the real part of the friction $\Tilde{\Gamma}^{hyd}(\omega)$ in Eq.~\eqref{eq:resp_sphere} converges for $\omega\to \infty$ to a constant value depending on the steady-state viscosity constants and relaxation times. This stems from our choice of exponential-oscillatory models for shear and volume viscosity, where the real part of the components scales with $\omega^{-4}$ and the imaginary part scales with $\omega^{-1}$. The friction in the frequency domain differs only marginally for frequency-dependent volume viscosity and for vanishing volume viscosity, i.e., for $\Tilde{\zeta} = 0$. For $\Tilde{\zeta} = \zeta_0$, the real part diverges. For $\Tilde{\zeta} = 0$, the friction has the same asymptotic behavior as for $\Tilde{\zeta}(\omega)$ in Eq.~\eqref{eq:rf_real_asympt}, with modified constants. This is visible in Fig.~\ref{fig:resp_fct_visc_freq_volume_visc}.
	\\\indent In the abscence of slip, i.e., for $b=0$, the real part scales as 
	\begin{align}
		\label{eq:rf_real_asympt2}
		\frac{\:\text{Re}\:\Tilde{\Gamma}^{hyd}(\omega)}{6\pi\eta_0a} \simeq
		\begin{cases} 
			\omega \rightarrow 0 & 1 + \frac{a\sqrt{ \rho_0/\eta_0}}{\sqrt{2}} \omega^{1/2}+ \mathcal{O}(\omega^{3/2}), \\
			\omega \rightarrow \infty & \\
			\text{if } \Tilde{\zeta} = \zeta_{0} & \frac{2a}{9}\Phi \frac{(\alpha_\infty)^2}{2\lambda_c}  \omega^{1/2},
			\\\text{otherwise} &  \frac{2a}{9} \Phi  \frac{ \alpha_{\infty}(\alpha_\infty + 2 \lambda_\infty)}{\lambda_\infty} \:\omega^{0}.
		\end{cases}
	\end{align}
	
	Therefore, the plateau value for $\omega \to \infty$ and frequency-dependent $\Tilde{\zeta}$ depends on whether the slip length is zero or not, but the long-time scaling for $\Tilde{\zeta} = \zeta_{0}$ is slip-independent.
	\\\indent For completeness, the imaginary part of the friction function, for $b\neq 0$ has the following asymptotic scaling
	\begin{align}
		\label{eq:rf_imag_asympt}
		\frac{\:\text{Im}\:\Tilde{\Gamma}^{hyd}(\omega)}{6\pi\eta_0a} \simeq
		\begin{cases} 
			\omega \rightarrow 0 & \frac{a(1+2\hat{b})^2\sqrt{\omega \rho_0/\eta_0}}{\sqrt{2} (1+3\hat{b})^2}\\ & + \mathcal{O}(\omega), \\
			\omega \rightarrow \infty & \\\text{if } \Tilde{\zeta} = \zeta_{0} & \frac{2}{9a} \Phi \frac{(\alpha_\infty)^2}{2\lambda_c}  \omega^{1/2},
			\\\text{otherwise} & - \frac{2a}{9} \Phi C_\infty\omega^{-1}.
		\end{cases}
	\end{align}
	The constant $C_\infty$ is given by
	\begin{eqnarray}
		C_\infty = 4 + \frac{2}{\hat{b}}- \frac{(\alpha_{\infty})^2}{(\lambda_{\infty})^2}.
	\end{eqnarray}
	Thus, we see that for vanishing volume viscosity, i.e., $\Tilde{\zeta}(\omega) \rightarrow 0 $ as $\omega \rightarrow \infty$, the imaginary part of the friction decays as $\sim \omega^{-1}$ for high frequencies, but diverges with the same scaling as the real part for $\Tilde{\zeta} = \zeta_{0}$.
	
	For the stick case ($b\rightarrow 0$), the imaginary part scales as
	\begin{align}
		\label{eq:rf_imag_asympt2}
		\frac{\:\text{Im}\:\Tilde{\Gamma}^{hyd}(\omega)}{6\pi\eta_0a} \simeq
		\begin{cases} 
			\omega \rightarrow 0 & -\frac{a\sqrt{\omega \rho_0/\eta_0}}{\sqrt{2}}\\ & + \mathcal{O}(\omega), \\
			\omega \rightarrow \infty & \\\text{if } \Tilde{\zeta} = \zeta_{0} & \frac{2}{9} \Phi \frac{(\alpha_\infty)^2 a}{2\lambda_c}  \omega^{1/2},
			\\\text{otherwise} & - \frac{2a}{9} \Phi \frac{\alpha_{\infty}(\alpha_{\infty} - 4\lambda_{\infty})}{(\lambda_{\infty})^2} \omega^{-1}.
		\end{cases}
	\end{align}
	
	\section{Derivation of the Hydrodynamic Tail}
	\label{app:hydro_tail}
	Applying the inverse Fourier transformation in Eq.~\eqref{eq:mom_sphere12} leads to the Boussinesq-Basset equation \cite{boussinesq1903theorie,chow1972effect,lesnicki2016molecular}. We decompose the total force in Eq.~\eqref{eq:mom_sphere12} into three forces
	\begin{align}
		\label{eq:forces_BB}
		\Tilde{F}_{i,1}^{sp}(\omega) &= -6\pi a \Tilde{\eta}(\omega) \Tilde{V}_i^{sp}(\omega), \\
		\Tilde{F}_{i,2}^{sp}(\omega) &= -6\pi a^2 \sqrt{\Tilde{\eta}(\omega) \rho_0 i \omega} \Tilde{V}_i^{sp}(\omega), \\
		\Tilde{F}_{i,3}^{sp}(\omega) &= -\frac{2}{3}\pi a^3 \rho_0 i \omega \Tilde{V}_i^{sp}(\omega).
	\end{align}
	Assuming a frequency-independent shear viscosity, i.e., $\Tilde{\eta}(\omega) = \eta_0$, it is easily seen that in the time-domain, the first component is given by $F_{i,1}^{sp}(t) = - 6\pi a \eta_0 V_i^{sp}(t)$, and the third component by $F_{i,3}^{sp}(t) = - \frac{2}{3}\pi \rho_0 a^3 \Dot{V}_i^{sp}(t)$. To derive the expression for the second component, we use $\Tilde{V}_i^{sp}(\omega) = \Dot{\Tilde{V}}_i^{sp}(\omega)/i \omega$, where the dot denotes the time derivative. For the expression $\Tilde{F}_{i,2}^{sp}(\omega) = -6\pi a^2 \sqrt{\frac{\eta_0\rho_0}{i \omega}} \Dot{\Tilde{V}}_i^{sp}(\omega) = -6\pi a^2 \sqrt{\eta_0\rho_0} \Tilde{g}(\omega)\Dot{\Tilde{V}}_i^{sp}(\omega)$, we can show that $\Tilde{g}(\omega)$ in the time domain is given by $g(t) = \frac{\Theta(t)}{\sqrt{\pi}}t^{-1/2}$
	\begin{align}
		\label{eq:wick_integral}
		\Tilde{g}(\omega) &= \frac{1}{\sqrt{\pi}}\int_0^\infty e^{-i\omega t}\frac{dt}{t^{1/2}}, \\
		\label{eq:wick_integral2}
		&= \frac{1}{\sqrt{\pi i \omega}} \int_0^{i \infty} e^{-s} \frac{ds}{s^{1/2}}, \\
		\label{eq:wick_integral3}
		&=  \frac{1}{\sqrt{\pi i \omega}} \int_0^{\infty} e^{-s} \frac{ds}{s^{1/2}}, \\
		\label{eq:wick_integral4}
		& = \frac{1}{\sqrt{i \omega}},
	\end{align}
	where we use a Wick rotation to arrive at Eq.~\eqref{eq:wick_integral3} from Eq.~\eqref{eq:wick_integral2}. Applying the convolution theorem in the inverse Fourier transformation of the second force component, the full force is given by the Boussinesq-Basset equation
	\begin{align}
		\label{eq:mom_sphere11}
		F_{i}^{sp}(t) =&- 6\pi a \eta V_i^{sp}(t) - 6a^2\sqrt{\pi\eta_0\rho_0}\int_{0}^t dt'\:\frac{\Dot{V}_i^{sp}(t')}{\sqrt{t-t'}} \\ \nonumber &- \frac{2}{3}\pi \rho_0 a^3 \Dot{V}_i^{sp}(t).
	\end{align}
	The first term in Eq.~\eqref{eq:mom_sphere11} is the steady Stokes drag. The third term is known as the added mass term, where $m_{0} = \frac{2}{3} \pi \rho_0 a^3$. It originates since the accelerating sphere in the unsteady flow must move or deflect some surrounding fluid volume. The second term, including a convolution integral, describes the sphere's history of motion, also known as the Basset history force. Applying a partial integration on this term, we obtain
	\begin{align}
		\label{eq:basset_force}
		F_{i,2}^{sp}(t) =&  3a^2\sqrt{\pi\eta_0\rho_0}\int_{0}^t dt'\:(t-t')^{-3/2 }V_i^{sp}(t') \\\nonumber & + V_i^{sp}(t)f(0)
		- V_i^{sp}(0)f(t),
	\end{align}
	where $f(t) = \theta(t) 6a^2\sqrt{\pi\eta_0\rho_0}t^{-1/2} $.
	For long times, we find that the memory kernel from the hydrodynamic prediction contains a power law decay
	\begin{equation}
		\label{eq:hydro_tail_app} 
		\lim \limits_{t \to \infty} \Gamma^{hyd}(t) \approx - 3a^2\sqrt{\pi\eta_0\rho_0}t^{-3/2},
	\end{equation}
	which is the famous hydrodynamic tail \cite{ernst1970asymptotic,corngold1972behavior, lesnicki2016molecular}.
	The Volterra equation in Eq.~\eqref{eq:G-Volterra3} relates the long-time behavior of the memory kernel and the velocity autocorrelation function $C^{vv}(t)$ (VACF). Using the expression in Eq.~\eqref{eq:c_vv_kernel_ft}, we find that for long times, the VACF scales with
	\begin{equation}
		\label{eq:hydro_tail_app_corrv} 
		\lim \limits_{t \to \infty} C^{vv}(t) \approx  \frac{k_BT \sqrt{\rho_0}}{12} (\pi\eta_0 t )^{-3/2}.
	\end{equation}
	
	\section{Calculation of the Frequency-Dependent Friction  from  Simulation Trajectories }
	\label{app:memkernel_iter}
	
	\begin{figure*}
		\centering
		\includegraphics[width=1
		\linewidth]{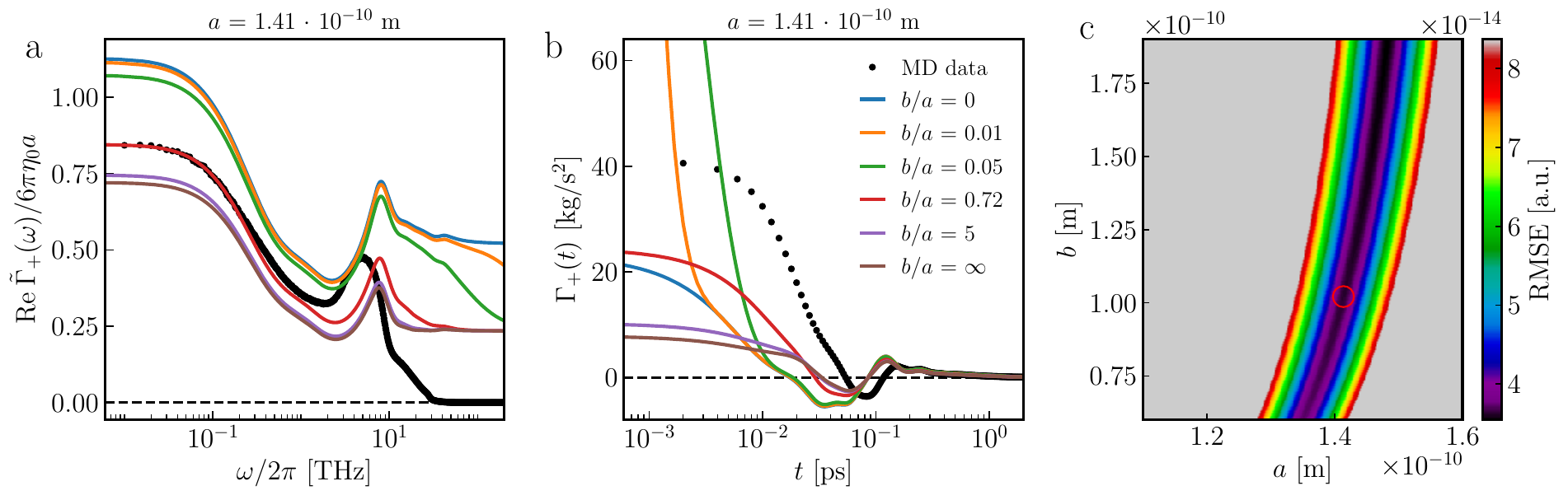}
		\caption{Influence of the slip length $b$ on the hydrodynamic friction. (a) Real part of the friction function Re$\:\Tilde{\Gamma}_+(\omega)$ in Fig.~\ref{fig:resp_fct_kernel_water_comparison} for different slip lengths $\hat{b} = b/a$, with fixed radius $a$ = 1.41 $\cdot\:10^{-10}$ m. (b)  Results of (a) in the time domain after numerical inverse Fourier transformation. (c) Root mean squared error (RMSE) between the MD data of the real part of the friction Re$\:\Tilde{\Gamma}_+(\omega)$  in Fig.~\ref{fig:resp_fct_kernel_water_comparison} (black), and the fit of the data using Eq.~\eqref{eq:resp_sphere} in dependence of the sphere radius $a$ and the slip length $b$. The red circle denotes the optimal parameter combination corresponding to the result in green in Fig.~\ref{fig:resp_fct_kernel_water_comparison} and shown as red lines in (a) and (b).}
		\label{fig:resp_fct_kernel_water_varied_radius_slip}
	\end{figure*}
	
	Various data-based methods to estimate the parameters of the GLE from experimental or simulation trajectories have been proposed \cite{lange2006collective, jung2017iterative, daldrop2018butane, klippenstein2021cross, vroylandt2021likelihood}.
	A robust and computationally efficient technique to extract the memory kernel from given time series trajectories can be derived by multiplying Eq.~\eqref{eq:GLE} for the position $x$ with the initial velocity  $\Dot{x}(0)$. Taking the ensemble average leads to an equation involving correlation functions that can be calculated from the given trajectory \cite{daldrop2018butane, berne1970calculation, straub1987calculation}. By this, we obtain a Volterra equation of the first kind \cite{kowalik2019memory, scalfi_pbc}
	\begin{equation}
		\label{eq:Volterra}
		mC^{\Dot{x}\Ddot{x}}(t) = - \int_0^t dt'\: \Gamma(t')C^{\Dot{x}\Dot{x}}(t-t'),
	\end{equation}
	where $C^{\Dot{x}\Ddot{x}}(t)$ = $\langle \Dot{x}(0)\Ddot{x}(t)\rangle$, and  $C^{\Dot{x}\Dot{x}}(t)$ = $\langle \Dot{x}(0)\Dot{x}(t)\rangle$, and we used that $\nabla U = 0$ and that $\Dot{x}(0)$ and $F_R$(t) are uncorrelated, i.e., $\langle \Dot{x}(0)F_R(t)\rangle$ = 0 \cite{mori1965transport}. \\
	\indent Analyses for one-dimensional trajectories have shown that compared to the direct method \cite{daldrop2018butane}, extraction of the memory kernel's running integral produces significantly more stable results \cite{kowalik2019memory}. We integrate Eq.~\eqref{eq:Volterra} over time
	\begin{align}
		\label{eq:G-Volterra1}
		m (C^{\Dot{x} \Dot{x}}(t) - C^{\Dot{x} \Dot{x}}(0)) &= - \int_0^t ds\:\int_0^s ds'\:\Gamma(s') C^{\Dot{x} \Dot{x}}(s-s'),\\
		&=  - \int_0^t ds'\int_{s'}^t ds\:\Gamma(s-s') C^{\Dot{x} \Dot{x}}(s'),\\
		&= - \int_0^t ds'\:\int_{0}^{t-s'} du \:\Gamma(u) C^{\Dot{x} \Dot{x}}(s'),\\
		\label{eq:G-Volterra3}
		&= - \int_0^t ds\:G(t-s) C^{\Dot{x} \Dot{x}}(s),
	\end{align}
	
	where $G(t)$ = $\int_0^t ds\:\Gamma(s)$ is the running integral of the memory kernel. Discretizing this equation with a time step $\Delta t$, we obtain an iterative formula for $G_i$ = $G(i\Delta t)$. For a discretized correlation function we use the short-hand notation $C_i^{AB}$ = $\langle A(0) B(i\Delta t) \rangle$. For the running integral of the memory kernel $G_i$, we obtain from Eq.~\eqref{eq:G-Volterra3} by applying the trapezoidal rule on the integral
	\begin{equation}
		\label{eq:G-iter}
		G_i  = \Bigl\lbrack  m( C_0^{\Dot{x}\Dot{x}} - C_i^{\Dot{x}\Dot{x}}) - \Delta t \sum_{j=1}^{i-1}G_j C_{i-j}^{\Dot{x}\Dot{x}}\Bigr\rbrack \cdot (\frac{1}{2}\Delta \:  t\:C_0^{\Dot{x}\Dot{x}})^{-1}, 
	\end{equation}
	where we use $G_0$ = 0. If we compute the velocity autocorrelation function $C_i^{\Dot{x}\Dot{x}}$ from the given time series $x(t)$, we can use Eq.~\eqref{eq:G-iter} to determine the running integral $G(t)$ and based on this the memory kernel $\Gamma(t)$ by differentiation.

	\section{Frequency-Dependent finite-size Correction for Memory Kernels}
	\label{app:pbc_correction}
	
	\begin{figure*}
		\centering
		\includegraphics[width=0.8
		\linewidth]{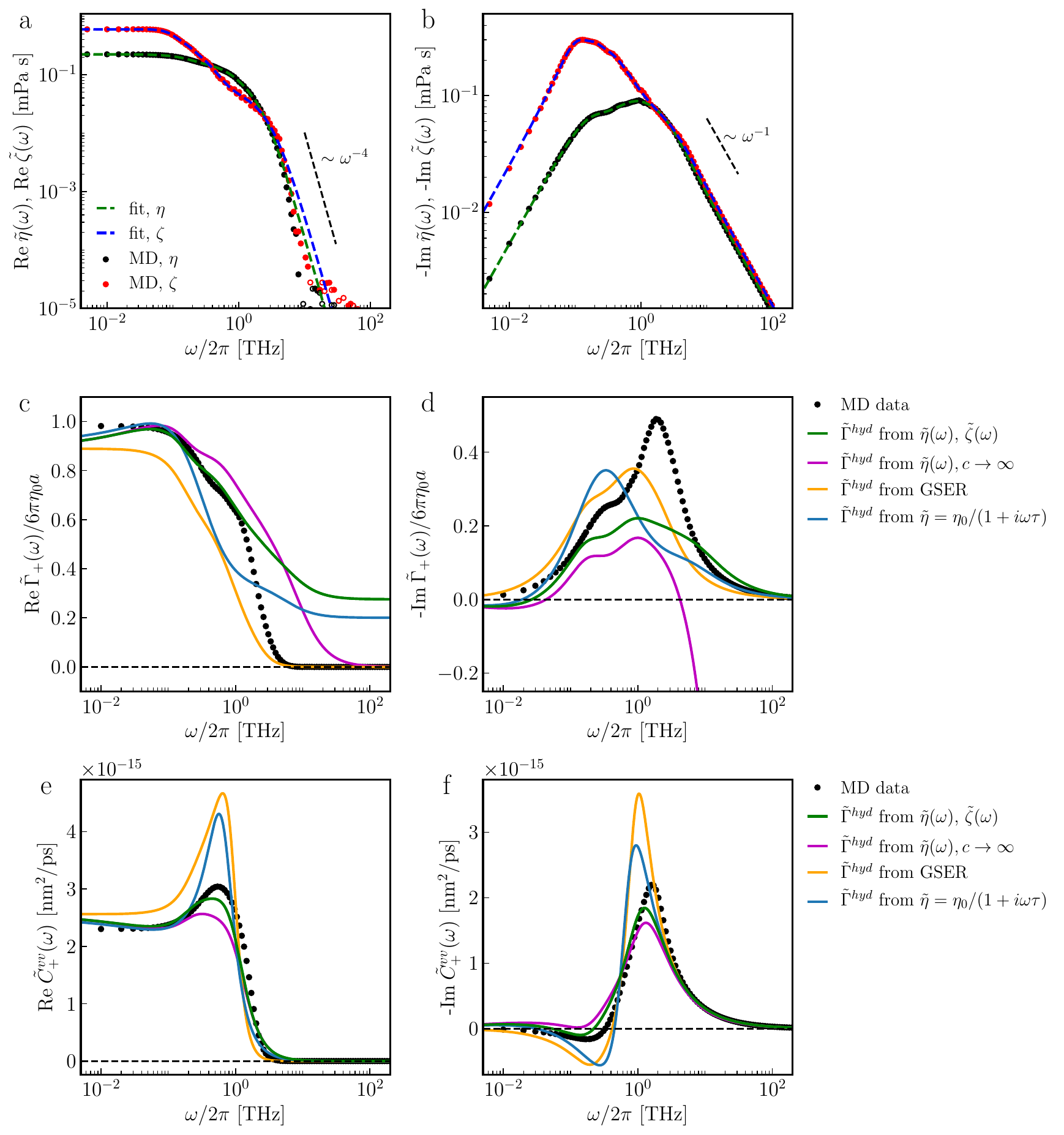}
		\caption{Simulation results and hydrodynamic theory for a Lennard-Jones fluid. (a, b) Extracted shear viscosity $\eta$ and volume viscosity $\zeta$ in frequency space from MD simulations (symbols) with fits (dashed lines) according to Eq.~\eqref{eq:visc_model} and the fitting parameters in Table~\ref{tab:fit_visc_lj} (Appendix \ref{app:fit_params}). Empty circles denote negative values. Note that all data is shown in logarithmic spacing for better visibility. (c, d) Real and imaginary part of the Fourier transformed friction $\Tilde{\Gamma}_+(\omega)$ 
			extracted from MD simulations using the GLE (symbols). We compare with the results from Eq.~\eqref{eq:resp_sphere} (green) for $a = 132.36$ pm and $\hat{b} = b/a = 0.17$ and the shear and volume viscosity parameters given in Appendix \ref{app:fit_params}. We also compare with the hydrodynamic prediction neglecting compressibility, i.e., $c \to \infty$ (purple), and additional neglected shear-wave effects, i.e., $\alpha \to 0$ (yellow, Eq.~\eqref{eq:gser}). The blue lines show the hydrodynamic friction using a Maxwell model fit of the shear viscosity to the MD friction data and vanishing volume viscosity. The fitting constants are $\eta_{0}$ = 0.23 mPa s and $(2\pi\cdot\tau)^{-1 }$ = 0.34 THz. (e, f) Comparison between the frequency-dependent velocity autocorrelation function $\Tilde{C}^{vv}_+(\omega)$ of a LJ particle in a LJ fluid from the MD and the hydrodynamic prediction (Eq.~\eqref{eq:c_vv_kernel_ft}), which we obtain using the results shown in (c, d).}
		\label{fig:lj_results}
	\end{figure*}
	
	Numerical results of the velocity autocorrelation function and the frequency-dependent friction of a particle in a
	fluid are significantly affected by the presence of periodic boundary conditions and finite system sizes in molecular dynamics simulations \cite{yeh2004system}. We follow the procedure elaborated in \cite{scalfi_pbc} which yields an analytic correction term for the frequency-dependent friction that accounts for periodic boundary effects. From the uncorrected friction $\tilde{\Gamma}_+(\omega)$, the corrected version is computed by
	\begin{equation}
		\label{eq:finite_size1}
		\bigl [ \tilde{\Gamma}_+^{\text{corr}}(\omega)  \bigr]^{-1} = \bigl [\tilde{\Gamma}_+(\omega) \bigr]^{-1} - \Delta G^{\text{corr}}(\omega),
	\end{equation}
	where the frequency-dependent correction term $\Delta G^{\text{corr}}(\omega)$ is given by
	\begin{align}
		\label{eq:finite_size2}
		\Delta G^{\text{corr}}(\omega) = \Bigl[ & \sum_{\vec{n},\Vec{n}\neq 0} \frac{1}{3} \text{Tr}[G_{ij}(\vec{n}L,\omega)]\Bigr]  \\\nonumber -& \frac{1}{3L^3} \int d\vec{r}' \text{Tr} [G_{ij}(\vec{r}',\omega)]
		.
	\end{align}
	$L$ is the box size, here $L = 3.5616$ nm for the water simulation and $L = 4.5$ nm for the methane in water simulation \cite{kowalik2019memory}, and $\vec{n} = n_x\vec{e}_x + n_y\vec{e}_y + n_z\vec{e}_z$ is a lattice vector with $n_x$, $n_y$, $n_z$ integers and $\vec{e}_i$
	are the unit vectors in the directions $x$, $y$, and $z$. Tr denotes the tensor trace. $G$ is the Green's function given in Eq.~\eqref{eq:Greensfct_app}, derived in Appendix \ref{app:friction_deriv}. For a detailed description of the numerical calculation of the correction term in Eq.~\eqref{eq:finite_size2}, we refer to \cite{scalfi_pbc}. We additionally correct the velocity autocorrelation function $\tilde{C}_+^{vv}(\omega)$ from the MD simulations by the correction term
	
	\begin{align}
		\label{eq:finite_size3}
		\tilde{C}_+^{vv,\text{corr}}(\omega) &= \\\nonumber &\frac{ \tilde{C}_+^{vv}(\omega)}{1+(k_BT)^{-1}\tilde{C}_+^{vv}(\omega)\tilde{\Gamma}_+^{\text{corr}}(\omega)\tilde{\Gamma}_+(\omega)\Delta\Tilde{G}^{\text{corr}}(\omega)},
	\end{align}
	which is derived using Eq.~\eqref{eq:c_vv_kernel_ft}.
	The corrected data $\tilde{\Gamma}_+^{\text{corr}}(\omega)$ and $\tilde{C}_+^{vv,\text{corr}}(\omega)$ is shown in the Figs.~\ref{fig:resp_fct_kernel_water_comparison}, \ref{fig:corrv_water_comparison} and \ref{fig:resp_fct_kernel_methane_comparison} as black circles. 
	Note that for the calculation of the PBC correction, we used the fitted frequency-dependent models shown in Fig.~\ref{fig:visc_water} as it improves the finite-size correction \cite{scalfi_pbc}.

	\section{Determination of the Effective Radius and the Slip Length}
	\label{app:slip_fit}
	
	In Fig.~\ref{fig:resp_fct_kernel_water_varied_radius_slip}, we investigate the dependence of the friction Re$\:\Tilde{\Gamma}_+(\omega)$ on the hydrodynamic radius $a$ and the slip length $b$. We observe a significant dependence of the slip length in Fig.~\ref{fig:resp_fct_kernel_water_varied_radius_slip}(a), where higher slip leads to lower friction. The same is visible in the time domain in Fig.~\ref{fig:resp_fct_kernel_water_varied_radius_slip}(b), obtained from (a) by applying the inverse Fourier transformation on $\Tilde{\Gamma}_+(\omega)$. 
	\\\indent The slip length seems to have no major influence on the shape but only on the magnitude of the friction function. 
	In Fig.~\ref{fig:resp_fct_kernel_water_varied_radius_slip}(c), we
	show the root mean squared error (RMSE) between the MD data of the real part of the friction Re$\:\Tilde{\Gamma}_+(\omega)$  (Fig.~\ref{fig:resp_fct_kernel_water_comparison}(b, black)), and the fit of the data using Eq.~\eqref{eq:resp_sphere} in dependence of the sphere radius $a$ and the slip length $b$. We can find a global minimum, denoted as a red circle, which corresponds to the results reported in Fig.~\ref{fig:resp_fct_kernel_water_comparison} and shown as a red line in Fig.~\ref{fig:resp_fct_kernel_water_varied_radius_slip}(a, b). 
	Due to anisotropic dependence of the error in Fig.~\ref{fig:resp_fct_kernel_water_varied_radius_slip}(c) on $a$ and $b$,
	we can fit the effective radius much more accurately than the slip length. Hence, the estimation of the slip length by this procedure is subject to considerable uncertainty.
	
	\section{Results for a Lennard-Jones Fluid}
	\label{app:lj_fluid}
	We here discuss simulation results and the hydrodynamic theory for a Lennard-Jones (LJ) particle in a LJ fluid (Appendix \ref{app:MD_sim} for simulation details). In Fig.~\ref{fig:lj_results}(a, b), we show the extracted and fitted viscoelastic spectra of the MD simulation. Note that, in contrast to the water system, we used four viscoelastic fitting components for both shear and volume viscosity. Also note that we do not subtract finite-size effects here, as we are only interested in the behavior in the intermediate frequency regime.  The extracted friction $\Tilde{\Gamma}_+(\omega)$ in Fig.~\ref{fig:lj_results}(c, d) agrees well with the hydrodynamic prediction using the fitted viscosity models (green), the agreement is worse with the model neglecting compressibility, i.e., $c \to \infty$ (purple), and the GSER (yellow). We used $\rho_0$ = 1370 kg/m$^3$ and $c$ = 869 m/s \cite{scalfi_pbc}. Even if no distinct peaks can be seen in the friction spectrum, several viscoelastic components are still necessary for the modeling, as a simple Maxwell model (blue) deviates significantly from the MD data in Fig.~\ref{fig:lj_results}(c, d). As in the water model, a plateau can be seen in the real part of the hydrodynamic friction for frequencies above 10 THz, which is not present in the MD data. Nevertheless, we find that the hydrodynamic prediction with the correct viscoelastic model and finite compressibility is the most appropriate one for the LJ fluid, which becomes even more evident when analyzing the VACF in Fig.~\ref{fig:lj_results}(e, f).

	\section{Long-Time Behavior of the Water Memory Kernel with Compressibility}
	\label{app:hydro_tail_compress}
\begin{figure}
		\centering
		\includegraphics[width=0.8
		\linewidth]{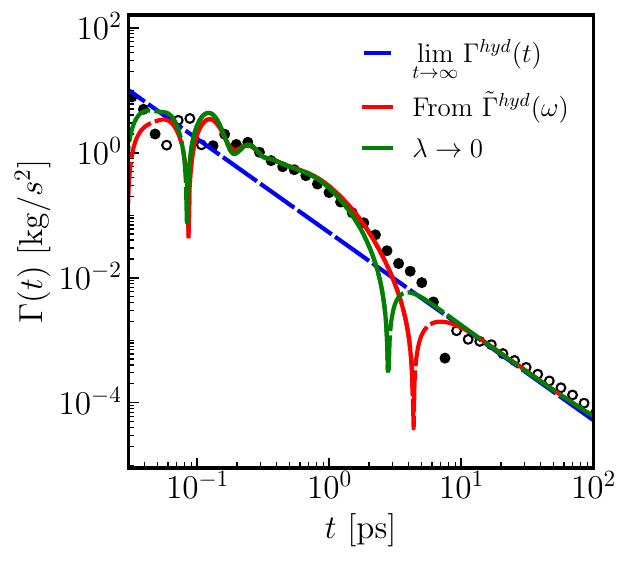}
		\caption{Discussion of the influence of compressibility on the hydrodynamic long-time tail. We compare the long-time behavior of the water memory kernel from MD simulations (symbols) with the numerical inverse Fourier transformation of the hydrodynamic friction $\Tilde{\Gamma}^{hyd}(\omega)$ in Eq.~\eqref{eq:resp_sphere} for the compressible case (red line) and in the incompressible limit, i.e., $\lambda \rightarrow 0$ (green line). Dashed lines and empty circles denote negative values.}
		\label{fig:resp_fct_kernel_hydro_tail_compress}
	\end{figure}

	In Fig.~\ref{fig:resp_fct_kernel_hydro_tail_compress}, we compare the long-time behavior of the water memory kernel from MD simulations (symbols) with the numerical inverse Fourier transformation of the hydrodynamic friction $\Gamma^{hyd}(\omega)$ in Eq.~\eqref{eq:resp_sphere} for the compressible case (red line) and in the incompressible limit, i.e., $\lambda \rightarrow 0$ (green line). As discussed in the main text, the hydrodynamic tail remains unchanged in the incompressible case. The sign change from positive to negative values around 3 ps is slightly shifted to earlier times.
	
	\section{Derivation of Eq.~\eqref{eq:c_vv_kernel_ft}}
	\label{app:deriv_cvv_ft}
	We start from the Fourier-transformed GLE in Eq.~\eqref{eq:GLE_FT}, i.e., $\Tilde{v}(\omega) = i \omega \Tilde{\chi}(\omega) \Tilde{F}_R(\omega)$, where $\Tilde{v}(\omega) = i\omega \Tilde{x}(\omega)$ and $\Tilde{\chi}(\omega) = (i \omega \Tilde{\Gamma}_+(\omega) - m\omega^2)^{-1}$. The fluctuation-dissipation theorem, i.e., $
	\langle F_R(t)F_R(t')\rangle = k_B T\:\Gamma(|t-t'|)$, in Fourier space reads
	\begin{align}
		\langle \Tilde{F}_R(\omega) \Tilde{F}_R(\omega')\rangle &= \\ \nonumber k_BT& \int_{-\infty}^\infty& dt\:e^{-i\omega t} \int_{-\infty}^\infty dt'\:e^{-i\omega' t'}  \Gamma(t-t'), 
	\end{align}
	\begin{align}
		= k_BT \int_{-\infty}^\infty dt'\:e^{-i(\omega+\omega') t} &\int_{-\infty}^\infty dt\:e^{-i\omega (t- t')}  \Gamma(t-t'), \\
		=&  k_BT \int_{-\infty}^\infty dt'\:e^{-i(\omega+\omega') t'}  \Tilde{\Gamma}(\omega), \\
		=& 2\pi k_BT \delta(\omega+\omega') \Tilde{\Gamma}(\omega).
	\end{align}
	Using this identity, we obtain the Fourier transformation of the VACF
	\begin{align}
		\Tilde{C}^{vv}(\omega)&= \int_{-\infty}^\infty \frac{d\omega'}{2\pi} e^{-i\omega (t-t)} \langle \Tilde{v}(\omega) \Tilde{v}(\omega')\rangle,\\
		&= \int_{-\infty}^\infty \frac{d\omega'}{2\pi} \langle i\omega \Tilde{x}(\omega) i \omega' \Tilde{x}(\omega')\rangle, \\
		&= - \int_{-\infty}^\infty \frac{d\omega'}{2\pi} \langle \omega \Tilde{\chi}(\omega)\Tilde{F}_R(\omega)\omega' \Tilde{\chi}(\omega')\Tilde{F}_R(\omega')\rangle, \\
		&= - k_BT \int_{-\infty}^\infty d\omega' \omega \omega' \Tilde{\chi}(\omega) \Tilde{\chi}(\omega') \delta(\omega+\omega') \Tilde{\Gamma}(\omega), \\
		&= k_BT \omega^2 \Tilde{\chi}(\omega) \Tilde{\chi}(-\omega) \Tilde{\Gamma}(\omega), \\
		&= k_BT \omega^2 \Tilde{\Gamma}(\omega) \frac{\Tilde{\chi}(-\omega) - \Tilde{\chi}(\omega)}{\frac{1}{\Tilde{\chi}(\omega)} - \frac{1}{\Tilde{\chi}(-\omega)}}, \\
		\label{eq:cvvft5}
		&= k_BT \omega^2 \Tilde{\Gamma}(\omega) \frac{\Tilde{\chi}(-\omega) - \Tilde{\chi}(\omega)}{i\omega (\Tilde{\Gamma}_+(\omega) + \Tilde{\Gamma}_+(-\omega))}.
	\end{align}
	Since $\chi(t)$ is a real function, we have $\Tilde{\chi}(-\omega) - \Tilde{\chi}(\omega) = \Tilde{\chi}^*(\omega) - \Tilde{\chi}(\omega) = - 2i \:\text{Im}\: \Tilde{\chi}(\omega)$, where $\Tilde{\chi}^*(\omega)$ is the complex conjugate of $\Tilde{\chi}(\omega)$. For any function $f(t)$ symmetric in $t$, such as $\Gamma(t)$ and $C^{vv}(t)$, we have
	\begin{align}
		\Tilde{f}_+(\omega) +  \Tilde{f}_+(-\omega) &= \Tilde{f}_+(\omega) +  \Tilde{f}_+^*(\omega), \\
		= \int_{-\infty}^\infty dt\:f_+(&t) e^{-i\omega t} + \int_{-\infty}^\infty dt\:f_+(t) e^{i\omega t}, \\
		= \int_{-\infty}^\infty dt\:f(t)& \theta(t) e^{-i\omega t} + \int_{-\infty}^\infty dt\:f(t) \theta(-t) e^{-i\omega t}, \\
		&= \Tilde{f}(\omega).
	\end{align}
	Inserting this identity for $\Tilde{\Gamma}_+(\omega)$ into Eq.~\eqref{eq:cvvft5}, we obtain
	\begin{align}
		\label{eq:cvvft62}
		\Tilde{C}^{vv}(\omega) = - 2 k_BT \omega \:\text{Im}\: \Tilde{\chi}(\omega),
	\end{align}
	and 
	\begin{align}
		\label{eq:cvvft7}
		\text{Re}\:\Tilde{C}_+^{vv}(\omega) &= - k_BT \omega \:\text{Im}\: \Tilde{\chi}(\omega), \\
		\label{eq:cvvft7b}
		&= k_BT \omega \:\text{Re}\:(i \Tilde{\chi}(\omega)),
	\end{align}
	where we used $\Tilde{C}^{vv}(\omega) = \Tilde{C}^{vv}_+(\omega) +  (\Tilde{C}^{vv}_+(\omega))^* = 2 \:\text{Re}\:\Tilde{C}^{vv}_+(\omega)$.
	For the single-sided VACF, we finally obtain
	\begin{align}
		\label{eq:cvvft8}
		\Tilde{C}_+^{vv}(\omega) &= i \omega k_BT  \Tilde{\chi}(\omega), \\ \nonumber &=  \frac{i \omega k_BT}{i \omega \Tilde{\Gamma}_+(\omega) - m \omega^2},
	\end{align}
	which is Eq.~\eqref{eq:c_vv_kernel_ft} in the main text.
	Here we use the fact, employing the Kramers-Kronig relations, that if the real parts of the Fourier transformations of two half-sided time-domain functions are equal (Eq.~\eqref{eq:cvvft7b}), the
	total complex functions are equal \cite{kowalik2019memory}. 
	\newpage
	\bibliographystyle{unsrt}
	\bibliography{refs}
\end{document}